\documentclass{article}

\usepackage{arxiv}

\usepackage[utf8]{inputenc} 
\usepackage[T1]{fontenc}    
\usepackage{hyperref}       
\usepackage{url}            
\usepackage{booktabs}       
\usepackage{amsfonts}       
\usepackage{nicefrac}       
\usepackage{microtype}      
\usepackage{lipsum}		
\usepackage{graphicx}
\usepackage{doi}

\usepackage{biblatex}
\usepackage{caption}
\usepackage{subcaption}
\usepackage{amsmath}
\usepackage{xcolor}
\usepackage{booktabs}
\usepackage{colortbl}
\usepackage[dvipsnames,svgnames,table]{xcolor}
\usepackage{xspace}
\usepackage{fontawesome5}  
\usepackage{hyperref}   

\DeclareMathOperator{\atantwo}{atan2}
\newcommand{\totalParticipants}{21\xspace}

\renewbibmacro*{name:andothers}{
  \ifboolexpr{
    test {\ifnumequal{\value{listcount}}{\value{liststop}}}
    and
    test \ifmorenames
  }
    {\ifnumgreater{\value{liststop}}{1}
       {\finalandcomma}
       {}%
     \andothersdelim\bibstring[\textit]{andothers}}
    {}}

\title{A Multi-Technique Approach for Improving Summary Polar Diagrams}

\author{ \href{https://orcid.org/0000-0002-0678-2870}{\includegraphics[scale=0.06]{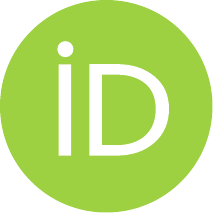}\hspace{1mm}Aleksandar Anžel}\thanks{\url{https://aanzel.github.io/}} \\
	Center for Artificial Intelligence in\\
    Public Health Research (ZKI-PH)\\
    Robert Koch-Institute\\
    Nordufer 20, Berlin\\
    13353, Germany\\
	\texttt{AnzelA@rki.de} \\
    \And
	\href{https://orcid.org/0000-0001-9974-3231}{\includegraphics[scale=0.06]{orcid.pdf}\hspace{1mm}Zewen Yang} \\
	Center for Artificial Intelligence in\\
    Public Health Research (ZKI-PH)\\
    Robert Koch-Institute\\
    Nordufer 20, Berlin\\
    13353, Germany\\
    \texttt{YangZ@rki.de} \\
	\And
	\href{https://orcid.org/0000-0003-4168-8254}{\includegraphics[scale=0.06]{orcid.pdf}\hspace{1mm}Georges Hattab}\thanks{\url{https://visualization.group/}} \\
	Center for Artificial Intelligence in\\
    Public Health Research (ZKI-PH)\\
    Robert Koch-Institute\\
    Nordufer 20, Berlin\\
    13353, Germany\\
	\texttt{HattabG@rki.de} \\
    \\
    Department of Mathematics and\\
    Computer Science, Freie Universität\\
    Arnimallee 14, Berlin\\
    14195, Germany\\
}

\renewcommand{\shorttitle}{Enhanced Summary Polar Diagrams}

\hypersetup{
pdftitle={A Multi-Technique Approach for Improving Summary Polar Diagrams},
pdfsubject={cs.OH},
pdfauthor={Aleksandar Anžel, Zewen Yang, Georges Hattab},
pdfkeywords={Polar coordinates, polar space, model comparison, overview detail, small multiple, taylor diagram, mutual information diagram.},
}

\begin{document}
\maketitle

\begin{abstract}
	While the polar system may lack the universal familiarity of its Cartesian counterpart, it remains indispensable for certain tasks. Summary polar diagrams, such as Taylor and mutual information diagrams, address tasks like discovering relationships, visualizing data similarity, and quantifying correspondence. Although these diagrams are invaluable tools for uncovering data relationships, their polar nature can hinder intuitiveness and lead to issues like overplotting. We present a hybrid approach that combines overview+detail, aggregation, interactive filtering, Cartesian linking, and small multiples to enhance the clarity, comprehensiveness, and functionality of summary polar diagrams. We performed a user study to assess this approach's effectiveness, noting comparable response times among participants. Additionally, three domain experts with varying visualization experience reviewed an implemented solution applying summary polar diagrams to climate, data science (novel), and machine learning, refining the approach prior to the user study. The findings underscore the versatility of our approach in enhancing comprehension, accessibility, and utility.
\end{abstract}


\section{Introduction}
\label{sec: introduction}

The selection of the coordinate system employed in data visualization can have an effect on the clarity and ease of interpretation of the displayed data. While the Cartesian system is popular for its familiarity, the polar system offers distinct advantages for data with angular or radial patterns. However, polar scatter charts like Taylor or mutual information diagrams face challenges like overplotting, where excessive data leads to clutter. This issue is amplified by the circular nature of polar charts. To address this in fields relying on these charts, effective strategies are needed to improve readability and interpretation.
The underutilization of polar coordinates in visualization is a complex issue. The lack of intuitive understanding may have hindered their adoption.
On the one hand, \textcite{polar-vs-cartesian-2} demonstrated that users analyzing polar visualizations often have slower response times when compared to their Cartesian counterparts, thus supporting these claims. 
On the other hand, \textcite{polar-vs-cartesian-1} showed that the use of polar (or radial) space does not induce the ``blinders effect'' and allows users to more easily discriminate and recall locations within such layouts.
Moreover, we could consider the lack of intuition not only as a cause but also as an effect~---~polar charts are not intuitive because of their underutilization. 
To break this \textit{circulus vitiosus}, we would either have to improve the intuitiveness of the charts or use them more in everyday tasks. 
The lack of use of polar coordinates in visualizing commonly found data does not improve the situation, even though these chart types are shown to be more appealing and engaging than their Cartesian counterparts~\cite{polar-appealing-1, polar-appealing-2, radial-linear-comparison}.

A significant number of polar charts such as chord diagram~\cite{intro_chord}, target diagram~\cite{target-diagram} and Taylor diagram  were developed for solving specific problems in specific domains. The data encountered in these domains is often naturally represented in a polar form (\textit{e.g.}, mathematics, astronomy, physics, climatology, \textit{etc.}).
Notable exceptions are the pie chart and its variations, such as the donut chart and gauge chart, which found their applications in modern user interfaces~(UIs) and visual storytelling~\cite{pie-chart}.
A key point to note is that in the literature, terms like ``charts'', ``plots'', ``visualizations'', and ``diagrams'' are used interchangeably for polar charts. However, we'll use ``diagrams'' for summary polar diagrams and ``polar charts'' elsewhere.

To make polar charts more intuitive to read and hence easier to understand, many modern higher-level visualization libraries and tools employ one or many interactive functionalities that aim to ease the exploration and understanding of the underlying data visualized in the polar space. However, the implementation of such functionalities for basic polar chart types, such as polar scatter, line, area, and bar charts, is often lacking or is underdeveloped.
Suppose we narrow our search space to the higher-level visualization libraries available in Python programming language. In that case, the lack of interactive features but also of support for polar charts becomes quite apparent. At the time of writing, Bokeh~\cite{bokeh-library} and Plotnine~\cite{plotnine-library} had only rudimentary support for polar charts, while Altair~\cite{altair-library} only supported pie charts. Matplotlib~\cite{matplotlib-library} and Seaborn~\cite{seaborn-library} support polar charts but borrow most interactive elements from the Cartesian space and do not adapt them to the polar space. Plotly~\cite{plotly-library} has the best support for polar charts; however, a smaller number of interactive elements, such as interactive multi-selection filtering, are borrowed from the Cartesian space and not adequately adapted.
Moreover, common visualization principles and guidelines, such as well known ``Visual Information-Seeking Mantra''~---~\textit{Overview first, zoom and filter, then details on demand}, by Shneiderman~\cite{overview-first} or small multiple technique, are often unemployed with polar charts.

Our work focuses on a specific type of polar chart known as summary polar diagram (\textit{i.e.}, Taylor and mutual information diagrams). We present the first implementation of an overview+detail technique combined with aggregation and Cartesian linking, as well as the first implementation of a small multiple technique for visualizing data dynamics in the realm of summary polar diagrams. Furthermore, we extend both through the integration of interactive filtering.

Our main hypothesis is that providing clear instructions and intuitive guidelines will enable non-experts to effectively interpret and solve the tasks that summary polar diagrams are designed to address (more in~\autoref{subsec: tasks}), without requiring prior expertise in data visualization.
The central advancement presented in this work is the transformation of summary polar diagrams into an interactive and task-driven visualization framework that combines multiple complementary interactive idioms. These idioms act in synchrony to enhance the interpretability, expressiveness, and analytical utility of the diagrams across different task categories (such as cross-sample comparison, identification of structural or temporal trends, and in-depth exploration of individual data attributes). This development establishes summary polar diagrams as dynamic analytical tools that actively support the exploration and reasoning processes they were originally conceived to address.
More specifically, this research has three distinct contributions:


\begin{itemize}
    \item \textbf{Enhancement of the summary polar diagrams through a combination of overview+detail, aggregation, interactive filtering, and Cartesian linking techniques.} To our knowledge, this work includes the first implementation of these techniques in the context of summary polar diagrams. Furthermore, our research indicates that the integration of these techniques has not been previously employed even in the broader context of general polar scatter charts.
    \item \textbf{Enhancement of data dynamics visualization feature in summary polar diagrams through the small multiple technique.} We extend the ideas described by \textcite{polar-diagrams} (referred to from this point on as ``the related study''~or~TRS) for visualizing data dynamics in summary polar diagrams by further extending the ability to visualize two versions or time points per sample.
    \item \textbf{User study and example applications.} We demonstrate the effectiveness of the first combination in this list (overview+\allowbreak detail, aggregation, interactive filtering, and Cartesian linking) by a user study and an expert review. Moreover, using both the first and the second combination of techniques presented above, we extend the application of summary polar diagrams through three case studies (longitudinal climate data analysis, sample analysis, and machine learning hyper-parameter tuning evaluation).
\end{itemize}

\begin{figure*}
    \centering
    \begin{subfigure}[t!]{0.2\linewidth}
        \centering
        \includegraphics[width=\linewidth]{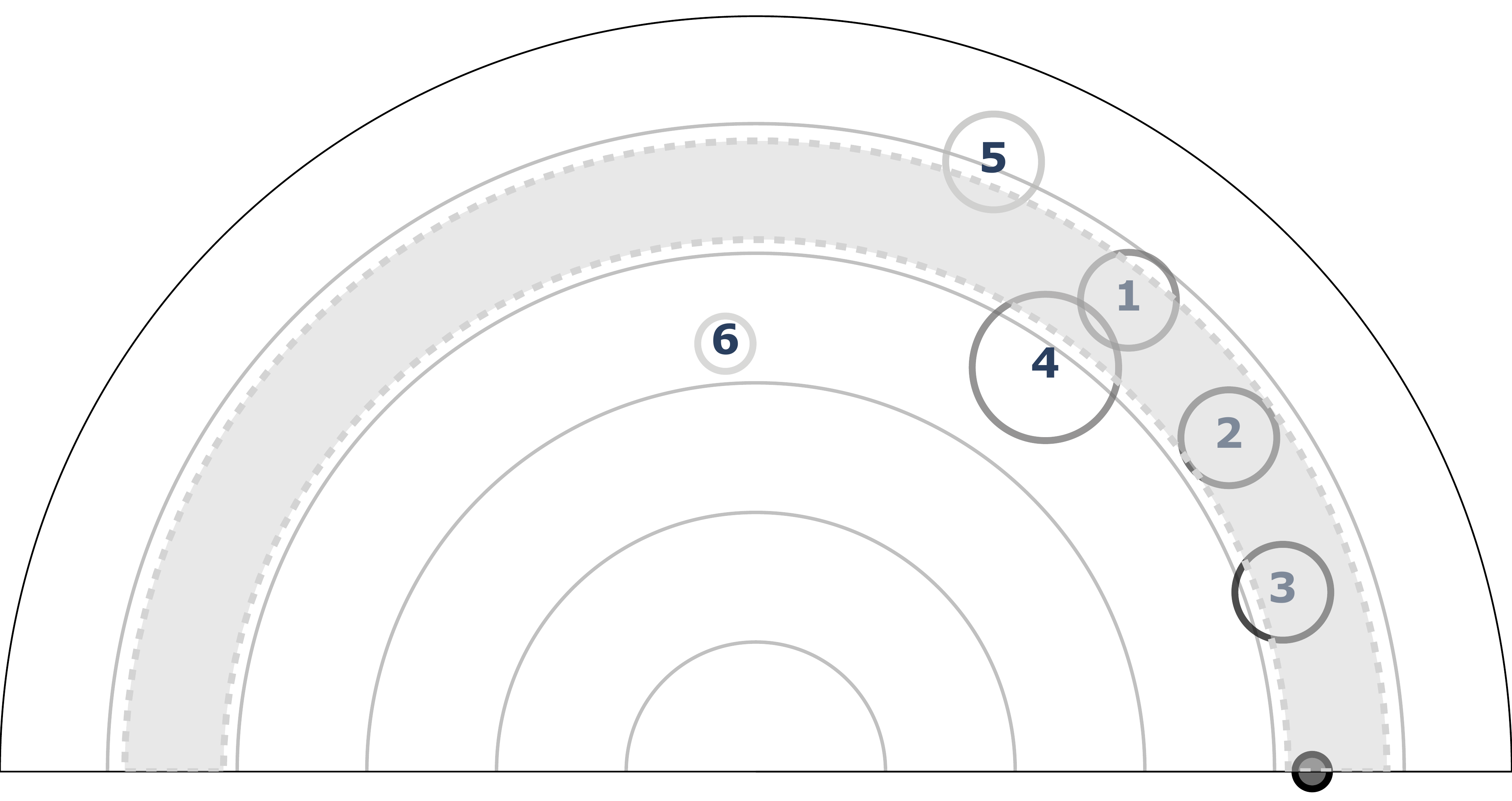}
        \includegraphics[width=0.5\linewidth]{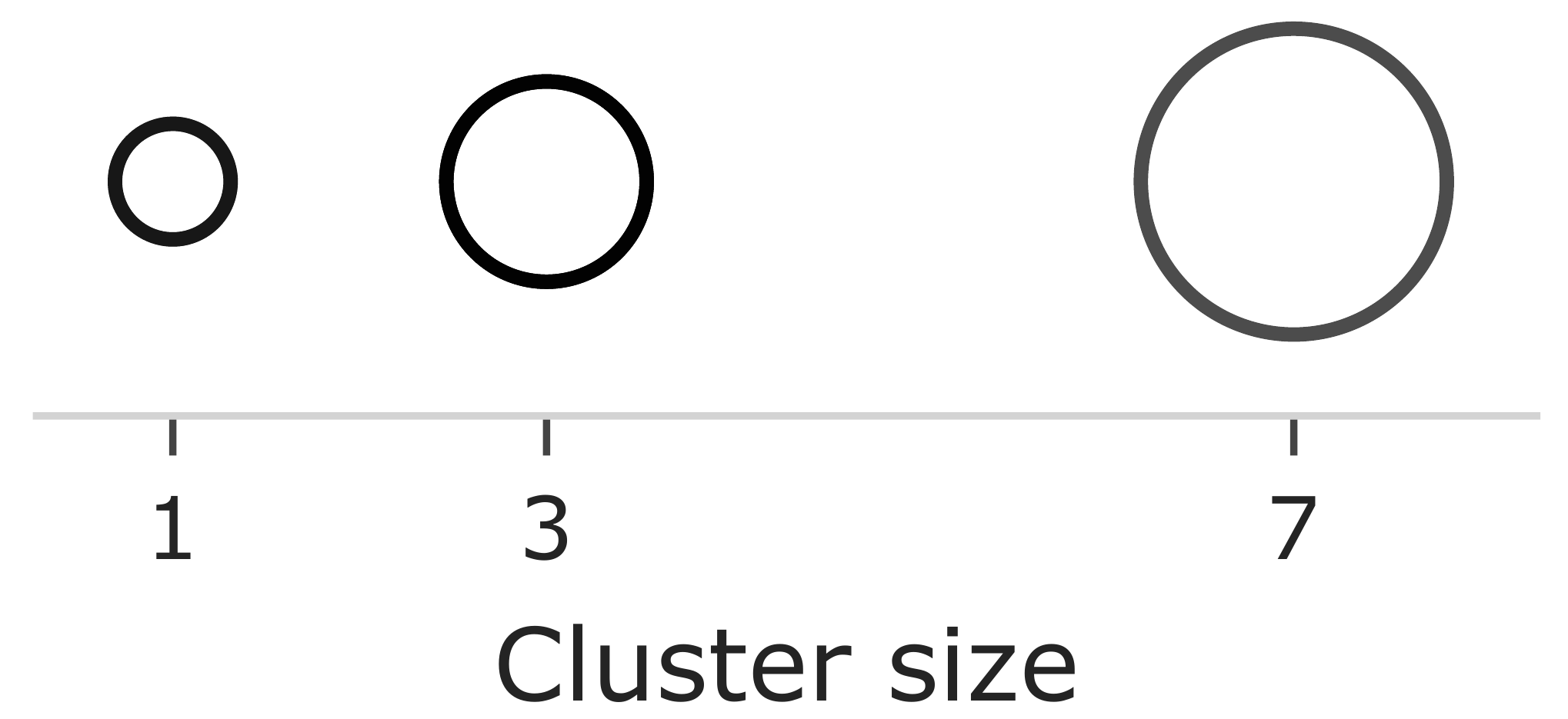}\\
        \vspace{0.5cm}
        \includegraphics[width=0.8\linewidth]{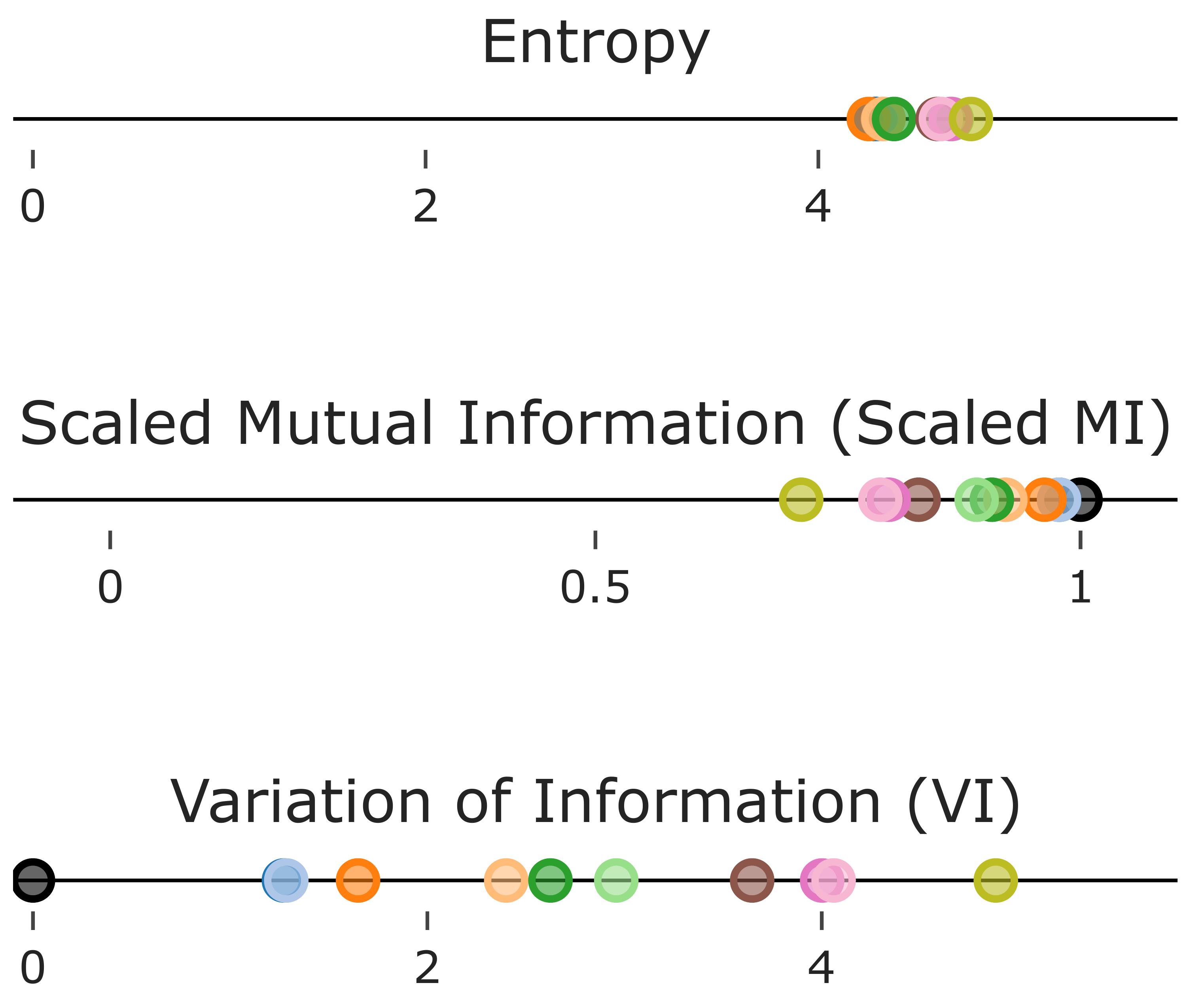}
        \label{fig: Wine_Overview}
    \end{subfigure}
    \hspace{1cm}
    \begin{subfigure}[t!]{0.7\linewidth}
        \centering
        \includegraphics[width=\linewidth]{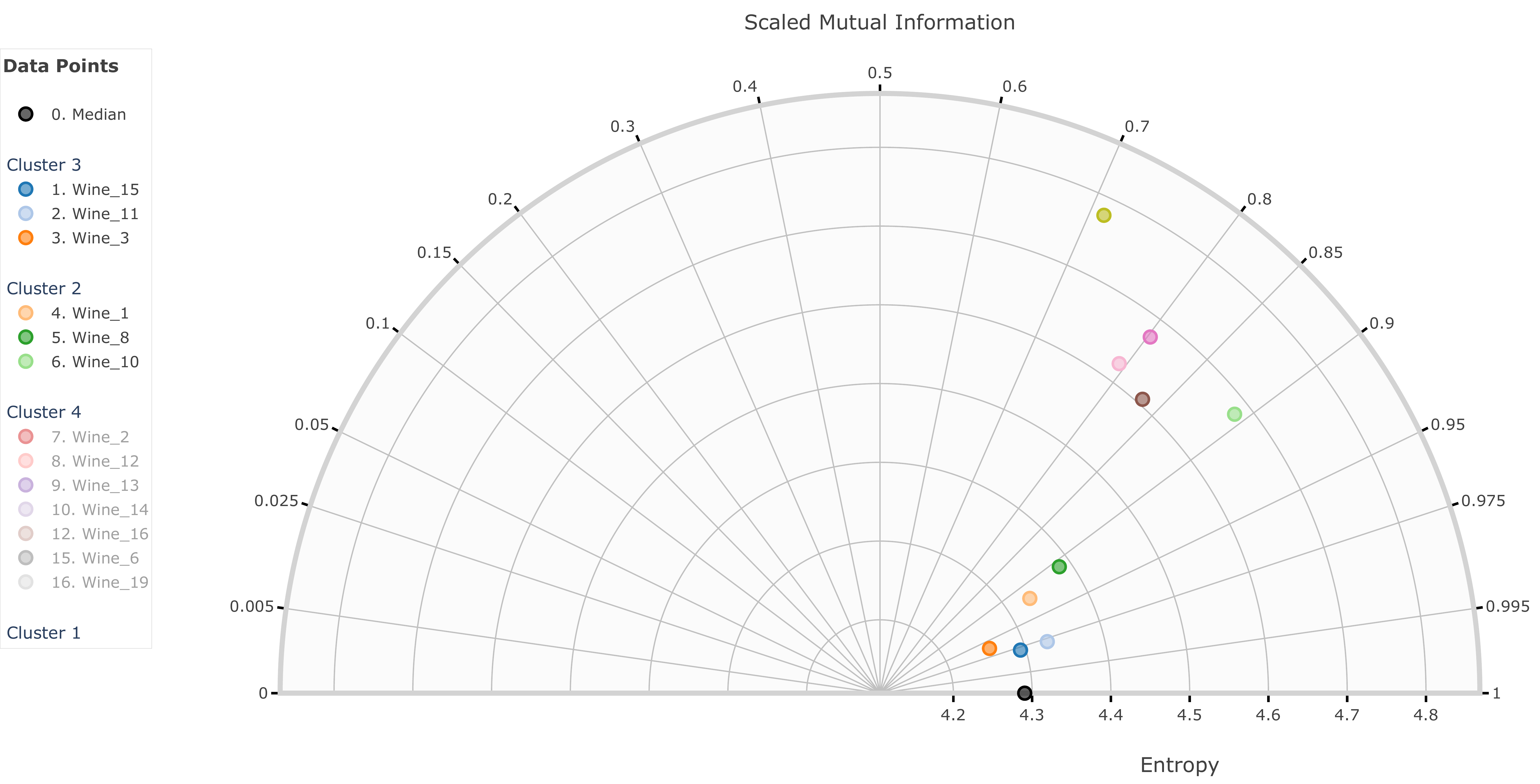}
        \label{fig: Wine_detail}
    \end{subfigure}
    \caption{\textbf{Enhanced scaled mutual information diagrams using overview+detail, aggregation, Cartesian linking, and interactive filtering}. The diagrams visualize nineteen wine samples compared to the theoretical wine sample containing median property values. The highlighted section in the overview indicates the region of interest, with the selected wine samples identified in the scrollable interactive legend, Cartesian-linking plot, and the detail diagram. The overview captures information about visual data clusters and their sizes, while the detail diagram and the Cartesian-linking plot encode exact information-theoretic values (entropy, scaled mutual information, and variation of information) of each sample.}
    \label{fig: Wine_comparison}
\end{figure*}

\section{Related Work}

This section offers a concise review of the most relevant themes from prior research on the application of clutter reduction and visualization techniques, including overview+detail, small multiples, and Cartesian linking, in the specific context of summary polar diagrams.

\subsection{Summary Polar Diagrams}
Instead of using $x$ and $y$ coordinates in a 2D Cartesian space, polar (or radial) visualization utilizes \textit{radial}~($r$) and \textit{angular}~($\theta$) axes of the polar coordinates space to encode the position of each individual data point. Converting between these two coordinate spaces involves basic trigonometry with the connection defined by~\autoref{eq: connection}.

\begin{equation}
\label{eq: connection}
    x = r\ \cos (\theta),\ y = r\ \sin (\theta),\ r = \sqrt{x^2 + y^2},\ \theta = \atantwo(y, x)
\end{equation}

Moreover, one could also use the position in the polar coordinate space to embed not only two but three different values if those values follow the cosine equation $c^2 = a^2 + b^2 - 2ab\cos{\theta}$.

This connection between the polar space and the cosine equation was first exploited by \textcite{taylor-diagram-original-paper} for the purposes of creating a new polar chart that embeds three summary statistics of two or more numerical vectors: standard deviation~($\sigma$), correlation~($R$), and centered root mean squared error~($CRMSE$). If we consider $(X, Y)$ as a pair of random variables, the connection between these three three summary statistics is depicted in~\autoref{equation-taylor}.

\begin{equation}\label{equation-taylor}
    CRMSE^2(X,Y) = \sigma_X^2 + \sigma_Y^2 - 2\sigma_X \sigma_Y R_{XY}
\end{equation}

Devised originally for comparing climate models, the Taylor diagram inspired researchers to explore summary polar and non-polar charts further.
Motivated to alleviate the shortcomings of the Taylor diagram such as the lack of the ability to detect non-linear relationships or the sensitivity to the outliers, \textcite{mutual-information-original-paper} presented the mutual information diagram. Recognizing the broader applicability of information-theoretic measures, which have since permeated a wide range of fields~\cite{information-theory-everywhere}, this polar diagram mitigates these issues by using such measures instead of the second order statistics to visualize each numerical vector. Aforementioned summary statistics are replaced with entropy~($H$), scaled mutual information~($SMI$), and variation of information~($VI$) in the case of scaled mutual information diagram. The authors also present a version of the diagram that only occupies the first quadrant of the polar space called the normalized mutual information diagram that instead utilizes root entropy~($\sqrt{H}$), normalized mutual information~($NMI$), and root variation of information~($RVI$). If we again consider $(X, Y)$ as a pair of random variables, the connection between the previously mentioned information-theoretic measures can be seen in~\autoref{equation-root-variation-of-information} in the case of normalized mutual information diagram, and~\autoref{equation-variation-of-information} in the case of scaled mutual information diagram.

\begin{equation}\label{equation-root-variation-of-information}
    \sqrt{VI(X,Y)}^2 = \sqrt{H(X)}^2 + \sqrt{H(Y)}^2 - 2\sqrt{H(X)} \sqrt{H(Y)} * NMI_{XY}
\end{equation}

\begin{equation}\label{equation-variation-of-information}
    VI^2(X,Y) = H^2(X) + H^2(Y) - 2H(X)H(Y) * (2 SMI_{XY} - 1)
\end{equation}

An improved implementation of both polar diagrams along with new features and interactivity, and the first public and open-source implementation of the mutual information diagram were presented in TRS.
In contrast, the work proposed here substantially advances beyond TRS by establishing a fully interactive and task-oriented framework for summary polar diagrams. Our implementation introduces multiple coordinated interactive idioms that were neither implemented nor conceptually addressed in TRS. These additions not only expand the functional interactivity of the diagrams but also markedly enhance their analytical power, enabling users to perform complex exploratory tasks with greater clarity, precision, and interpretive depth.
Additional diagrams inspired by the Taylor diagram and aimed at addressing its limitations include the solar diagram~\cite{solar-diagram} and the target diagram~\cite{target-diagram}. Nevertheless, the creators of these diagrams opted for a Cartesian coordinate system instead of a polar one to fulfill their objectives, thus placing these diagrams beyond the scope of this study.

\subsection{Overview+Detail, Small Multiple, and Cartesian-linking}

Overview+detail is a visualization technique employed to provide users with a comprehensive understanding of data while enabling them to concentrate on specific details. This technique integrates an overarching view with the option to zoom in on specific elements or regions for a detailed inspection. By offering an overview of the data in conjunction with the opportunity to delve into specific areas, users can effectively acquire insights.
The technique was previously employed to solve various exploratory tasks, including but not limited to reading electronic documents~\cite{overview-plus-detail-electronic-documents}, map navigation~\cite{overview-plus-detail-mobile, overview-plus-detail-maps}, topological analysis of large trees~\cite{overview-plus-detail-trees}, and web page exploration~\cite{overview-plus-detail-webpage}.
Despite its advantages (\textit{e.g.}, efficient navigation, improved spatial awareness, and providing a sense of control), multiple reviews of the overview+detail technique showed inconclusive results regarding its superior performance compared to other strategies~\cite{overview-plus-detail-mobile, overview-plus-detail-review-1, overview-plus-detail-review-2}. The performance of overview+detail was found to be highly dependent on the task at hand. 
However, notwithstanding its disadvantages (\textit{e.g.}, the additional screen space requirement and memory strain due to induced visual separation), overview+detail was almost always the preferred choice when users were presented with other strategies, even in studies where overview+detail performed poorly~\cite{overview-plus-detail-mobile}.

The technique of arranging multiple images in a grid to facilitate comparison existed long before the term \textit{small multiple} was coined by Tufte~\cite{small-multiple-tufte-origin} in 1990. Small multiple technique, also known as trellis displays or collections~\cite{small-multiple-large-singles}, can be defined in data science terms as a way to visualize a change (or dynamics) of a set of features over a fixed (usually larger) set of features of the same data set. Consequently, small multiple is often utilized instead of animation, where closer inspection of the static frames is paramount.
Just like the overview+detail technique, the efficacy of small multiple is dependent on the nature of the task at hand for the user. However, research findings consistently demonstrate that, for most tasks, small multiple outperforms animation, even though users often find animations more visually appealing~\cite{small-multiple-better-than-animation, small-multiple-better-than-animation-scatter, small-multiple-better-than-animation-dynamic}.

According to our research, no previous work has explored the idea of connecting visually or interactively scatter polar charts with their axes projection, which we here call the Cartesian-linking plot. Previous works have mostly focused on comparing and evaluating the effectiveness and usability of visualizations in polar and Cartesian coordinate spaces~\cite{polar-vs-cartesian-1, polar-vs-cartesian-2, polar-vs-cartesian-3, polar-vs-cartesian-4}. However, the potential benefits of integrating these two coordinate systems have not been thoroughly investigated. Our proposed Cartesian-linking plot aims to leverage the strengths of both polar and Cartesian visualizations, providing users with a more intuitive and informative way to explore data. Finally, while the aforementioned strategies have been utilized across a range of chart types, our research did not find any instances of them being applied in combination to summary polar diagrams. 

\subsection{Clutter Reduction through Aggregation and Interactive Filtering}

Reducing data complexity is a key strategy for managing clutter in visualizations. This can be accomplished, for example, through data aggregation, which groups data using a new representation, or data filtering, which excludes specific data from the visualization~\cite{munzner-book}. Data reduction strategy is frequently combined with additional techniques (\textit{e.g.}, overview+detail, small multiple) for improving the quality of the visualizations and reducing its complexity.

One of the most important aspects of clutter reduction is clutter detection and quantification. Previous works on the topic examined how to properly quantify visual clutter and proposed ideas for alleviating it~\cite{clutter-reduction-measures}. In addition, plenty of clutter reduction research has focused on scatter polar charts. \textcite{clutter-reduction-scatter-plot-improvement-sunspot-plots} created Sunspot Plots, which use a specialized function to smoothly transition between a discrete encoding in sparse regions and a density-based surface encoding in data-rich regions. Additionally, some works also define new ways to quantify clutter besides providing the means to reduce it~\cite{clutter-reduction-scatter-plot-automatic-with-metrics, clutter-reduction-scatter-plots-pixel-based-mappings}.

A taxonomy of clutter reduction techniques and a set of criteria that compares them was presented by \textcite{clutter-reduction-taxonomy}. Their research showed that no perfect technique exists, with all of them having strengths and weaknesses depending on the task and data at hand. In the specific case of clustering, their research showed the technique to be highly scalable but weak when there is a need to show attributes unless those attributes can be reduced to a summary statistic. In the specific case of summary polar diagrams, this weakness does not exist due to the nature of these diagrams. Moreover, they showcased that through the combination of multiple techniques, weaknesses can be alleviated.

Our literature review did not uncover any existing research on clutter reduction techniques specifically for polar charts. This gap in the utilization of this but also previously mentioned strategies within the context of summary polar diagrams presents an intriguing opportunity for exploring new approaches to data visualization in this particular domain. Furthermore, this work lays the groundwork for generalizing these ideas to basic polar scatter charts beyond summary polar diagrams.


\section{Enhanced Summary Polar Diagrams}
The enhanced summary polar diagrams present an extension of work started by TRS with the idea of improving readability, comprehension, and function of summary polar diagrams through a combination of various visualization techniques. To facilitate evaluation and user access, the original idea was transformed into a dashboard, incorporating a selection of data sets to explore the methods' potential. The dashboard defines a two-dimensional design space and a set of interaction elements.
Moreover, the implemented solution utilizes the \texttt{polar-diagrams} Python~\cite{python-book} library developed in TRS, along with the Dash framework~\cite{plotly-library}. This ensures the successful integration and implementation of all proposed visualization techniques within the interface.
For historical and conventional reasons, the following sections use the term ``model'' to refer to a data point within any summary polar diagram.

\subsection{Basic Ideas}

As highlighted in TRS, one of the hard constraints of the summary polar diagrams is the number of models these diagrams can distinguish. This limitation is a result of the qualitative aspect being encoded using the color channel, restricting the number of models that can be portrayed on the diagram. The current restriction allows for a total of twenty-one models~---~the reference model plus twenty additional models.

However, even with the number not being very high, models representing data from different fields can have quite similar statistical or information-theoretic properties. This similarity often leads to their placement close to each other on both the Taylor and mutual information diagrams.
The issue of visual mark occlusion representing each model was addressed in TRS through the implementation of three distinct strategies.
First, each mark color is semi-transparent, thus improving the readability of the overlapping marks. Second, the user is informed via a warning about the occlusion effect and provided with details on the marks affected by it. Third and last, among other interaction elements the diagrams include the semantic zooming which helps to tackle the aforementioned situation.

While the first two strategies effectively addressed the immediate challenge, the third strategy introduced a new concern~---~``out of sight, out of mind''. When zooming into the zones of interest, users risk losing the overall view of the entire polar space, potentially leading to information loss.
The integration of overview+detail and aggregation, along with semantic zooming alleviates this situation by always presenting juxtaposed overview and detail summary polar diagrams.

Even though the previous ideas address clutter reduction and potential information loss during zooming, they do not directly improve the analytical capabilities of the summary polar diagrams. These diagrams were originally designed to enable easy comparison and presentation of multiple models across different measures simultaneously. However, this can lead to mental saturation, especially for users unfamiliar with the polar coordinate space. To mitigate this issue, we introduced the ability to link each summary polar diagram to a corresponding 1-dimensional axis in the Cartesian coordinate system by juxtaposing the three measures encoded within the diagram. This element of the dashboard is referred to as the Cartesian-linking plot since it links the polar and Cartesian coordinate systems by projecting data from the former to the latter.

Additionally, as noted in TRS, the newly introduced feature of visually encoding multiple different versions of the same model simultaneously on summary polar diagrams was also subject to a constraint on the number of versions that could be displayed. In order to maintain visual clarity and avoid clutter, the authors imposed a limit of encoding only two different versions at the same time.
This limitation was addressed through the implementation of small multiple, which presents $n-1$ summary polar diagrams for $n$ different model versions.

\subsection{Design Space}
\label{subsec: design space}

The enhanced summary polar diagrams dashboard comprises a header that includes the description of the main enhancement technique (overview+detail or small multiple) in use, the data set name, and a selection box featuring three summary polar diagrams as options: Taylor, scaled mutual information diagram, and normalized mutual information diagram.
Additionally, the dashboard includes a navigation button positioned at the top left part of the header, which opens a sidebar containing supplementary information about the data sets and the data set selection tool. There are three options available: two for the overview+detail technique and one for the small multiple technique.

When presented with the \textbf{overview+detail enhancement technique}, the design space is split according to ``the rule of thirds''. The first third of the horizontal space is filled with the summary polar diagram, considered as an overview, while the other two-thirds are filled with the diagram containing the detailed view of the selected data set.
Immediately below the overview diagram lies a static legend depicting the scale of the mark sizes of the diagram. Another key component is the Cartesian-linking plot situated below the static legend. This plot features three 1-dimensional axes arranged vertically, with each axis corresponding to one of the measures encoded in the detail summary polar diagram. For the Taylor diagram, these axes represent standard deviation, correlation, and CRMSE. In the case of scaled (or normalized) mutual information diagram, the axes depict entropy (or its root), scaled (or normalized) mutual information, and variation of information (or its root).
Depending on the existence of occlusion effects from the marks in the detail diagram, a warning box can appear immediately below it. At the border of the first and the second third, and right between the two diagrams, an interactive legend lies, connecting them visually and in an information-specific sense, as it contains the combined information of both diagrams. The legend additionally connects the diagrams with the Cartesian-linking plot by showcasing the same models.

The overview diagram presents a visually simplified and abstracted version of the detail diagram. Besides removing the axis ticks, axis labels, and radial grid lines, the data representation is also deviated (further \textbf{aggregated}) from the representation used in the summary polar diagrams so far. Before being placed in the diagram, the data is first aggregated to create a visual abstraction in the overview. The statistical and information-theoretic model data is clustered using Density-Based Spatial Clustering of Applications with Noise (DBSCAN) algorithm~\cite{dbscan}, placing similar models in the same cluster. Each cluster is encoded using a circular mark with a solid border, transparent interior with text annotation, and three visual channels. First, the position channel encodes the average statistical or information-theoretic values of all models within the cluster. Second, the area channel reflects the number of models in the cluster~---~larger circles represent clusters with a greater number of models. Third, the color channel indicates the proximity to the reference model located on the radial axis~---~clusters closer to the reference model have darker borders on a black-and-white color scale. The text annotation within the mark interior indicates a cluster identifier, which is also present in the legend. This facilitates easier identification and interaction with clusters and their corresponding models.

The purpose of this process is threefold. First, it removes the visual occlusion of the marks that can occur in the smaller visual space of the overview diagram. Second, it allows users to easily exclude or include certain model clusters using the interactive legend connecting the two diagrams. Third, it improves the comprehension of data in the polar space by clustering models that are similar but do not visually appear close to each other in the polar space. One such example can be seen in~\autoref{subsec: wine}.
As for the detail diagram, we have kept its appearance consistent with the description provided in TRS.

The \textbf{Cartesian-linking} plot further enhances the readability of the detail summary polar diagram by unfolding its 2-dimensional polar space into three separate 1-dimensional Cartesian axes. The core purpose of summary polar diagrams is to enable the simultaneous comparison of models across multiple measures. However, this task can be mentally taxing, especially for users unfamiliar with these polar chart types.
To lower the barrier of adoption for these diagrams, the Cartesian-linking plot provides a complementary view that allows for closer, side-by-side comparison of the models across the different measures represented on the individual Cartesian axes. This helps users more easily discern the relationships between the various model attributes depicted in the detail summary polar diagram. All visual elements can also be seen in Supplemental Video 1.

The rule of thirds is also employed in the segmentation of the design space of the \textbf{small multiple enhancement technique}. The grid resolution of the design space is $ \lceil \frac{n-1}{3} \rceil \times 3$, with $n$ representing the number of versions of each model. Each summary polar diagram in a grid is annotated in the upper-right corner with the version number and the parameters that dynamically change across the grid. A shared interactive legend is located below the bottom-left diagram in the grid, facilitating cohesive data exploration across all diagrams.

\subsection{Interactions}

All diagrams presented via the dashboard using both overview+detail and small multiple techniques can be easily exported as static \texttt{SVG} images by using the ``Download plot'' button that appears upon hovering at the top-right position of the diagram. While this may seem like a minor detail, none of the previous implementations of summary polar diagrams have provided the ability to export them as vector image files. This ability was first enabled in TRS, and we have carried it forward in the enhanced implementation of summary polar diagrams presented here. Besides being informative even as static visualizations, the dashboard provides a set of user interactions that further enhance the fusion of the diagrams.

As mentioned in the previous section, the \textbf{overview+detail technique} employs two summary polar diagrams. The overview diagram allows for brushing interaction, enabling the selection of a desired polar (angular) sector highlighted in a semi-transparent gray color. The radial boundaries of the selection are visually linked to the detail diagram, which, upon brushing, is semantically zoomed (\textbf{filtered}) into the selected region. The selection also triggers two additional changes. First, the color of the axes and the background in the detail diagram changes to the highlighted gray color. Second, the brushing action highlights the corresponding marks in the main legend that connects the overview and detail diagrams, as well as in the Cartesian-linking plot. The legend offers both single and multi-selection options. Users can single-click or double-click on legend items to make their selections. Double-clicking on a model (mark) in the legend allows users to select and highlight (\textbf{filter}) the entire cluster to which the clicked model belongs. All interactions performed within the main legend are immediately reflected in both the detail diagram and the Cartesian-linking plot.
In line with TRS, the detail diagram includes a tooltip that is displayed when hovering over a mark. Furthermore, it implements a rectangular brush (\textbf{filter}) to enable the multi-selection of marks or regions of interest within the detailed view.
To facilitate easy model comparison, the Cartesian-linking plot includes a simple tooltip that displays the exact value for each model attribute represented on the 1-dimensional axes.

In the case of the \textbf{small multiple technique}, double-clicking on a mark in the legend enables users to select and highlight that mark across all diagrams in the grid. On the other hand, a single click action selects and highlights all marks except the clicked one. Furthermore, each diagram includes a rectangular brush feature, providing the ability to select multiple marks or regions of interest within the diagrams.
All interactions are also depicted in Supplemental Video 1.


\section{Expert Review and Pretest}
\label{sec: expert review}

In addition to the large-scale user study that gathered mostly quantitative structured data, we also performed an expert review with three professionals from diverse fields. This smaller-scale study served two purposes: first, to gather empirical qualitative data about our proposed hybrid approach, and second, to act as a pretest for our main user study, allowing us to refine both the study methodology and our proposed ideas.

\subsection{Participants}
We conducted a total of three reviews with experts~(E1~--~E3) from diverse fields:

\begin{itemize}
    \item \textbf{Expert 1}: A visualization professional and educator with extensive prior knowledge of summary polar diagrams.
    \item \textbf{Expert 2}: A bioinformatician and computer scientist with experience working with various types of data and limited prior knowledge of summary polar diagrams.
    \item \textbf{Expert 3}: A linguist with limited data science knowledge and no prior experience with summary polar diagrams.
\end{itemize}

These experts were all working professionals from distinctly different fields, with varying levels of data science knowledge and experience with summary polar diagrams. They all participated voluntarily in the study.

\subsection{Procedure}
This pretest study was conducted in the same manner as the main study, with the only difference being the use of an initial draft of the questionnaire. Some tasks from the initial draft remained unchanged in the final questionnaire, while others were slightly modified based on the feedback received during the pretest. The experts were informed that their task responses were not the primary focus; instead, they were encouraged to provide textual feedback on the questionnaire elements and the dashboard. Although we collected responses for all tasks in the pretest, we retained only the qualitative results from each expert.

\subsection{Results}
All experts reported that the dashboard's interface elements were well-integrated and appropriately positioned within an intuitive and user-friendly layout. The visual hierarchy and information architecture effectively guided users through the interface. Furthermore, each expert provided substantial feedback on the following components: data visualizations, interactive features, and overall system usability. We present a detailed analysis of their findings in the upcoming sections, which will elucidate specific areas of strength and potential improvements in the dashboard's design and functionality.

\subsubsection{Data Visualizations}
Several experts noted the ease of readability of all visual elements and the clear information they conveyed. In particular, E3 remarked that ``\textit{[the tutorial] clearly explained the scaled mutual information diagram, its elements, and how to read it}''. Moreover, the implementation of the Cartesian-Linking plot was well-received, as E2 noted that the ``\textit{[Cartesian-Linking] plot was quite helpful for solving the last task [Advanced information theoretic analysis, see Section 4 User Study in the main manuscript]}''. Finally, E1 highlighted the lack of uniformity between axis labels in the detail diagram and the Cartesian-Linking plot. We utilized this feedback to modify the labels and ensure consistency across both charts.

\subsubsection{Interactive Features}
Among interactive features, brushing presented the biggest challenge to all experts, with E3 and E2 noting that ``\textit{[...] brushing interaction is not intuitive}'' and ``\textit{[...] it is not clear how to reset the view after brushing}'', respectively. Similarly, E1 remarked on the lack of ability to select a polar sector that does not span the whole upper part of the circle ($0^{\circ} - 180^{\circ}$): ``\textit{I was expecting to be able to select a specific angular range as well, and not only radial}''.

The solutions to the first two remarks were quickly integrated as part of the tutorial, where we improved the screenshot depicting the brushing interaction and the accompanying text. E1's remark was particularly interesting because it highlighted the discrepancy in interactive features between Cartesian and polar coordinate charts. Since such a feature does not exist in any of the high-level visualization libraries and due to the significant time and effort required to implement it, we decided to postpone its development until we could test the usability of the existing version.


\subsubsection{System Usability}
All experts agreed on two issues regarding system usability: slow initial load time and significant lag between user interactions and system responses. The former was encountered during the dashboard loading process, while the latter occurred during brushing and multi-select interactive actions. The observed latency and sluggishness were primarily due to the dashboard being hosted on a user-shared, non-paid platform with limited resources. In contrast, these issues were absent when the tool was run locally or on a paid platform with dedicated resources. Performance was significantly improved by reallocating system resources, as evidenced by an upgrade from 512MB of random access memory with a 10\% CPU utilization cap for the entire hosting environment to the same memory allocation with a 50\% utilization threshold. Finally, E2 remarked on the interesting approach for analyzing data and expressed interest in ``\textit{[...] including it as a step in exploratory data analysis}''.

\section{User Study}
\label{sec: user study}

We conducted an experiment to evaluate the user experience of the enhanced summary polar diagrams that included overview+detail, aggregation, filtering, and Cartesian linking. Our research indicated that the combination of these techniques in the context of summary polar diagrams had not been explored before. Moreover, these diagrams had not been subjected to empirical evaluations in any previous reports. Motivated by the unexplored nature of this hybrid approach, we determined that an empirical evaluation of its performance and usability for our target end-users would be a valuable addition to the field. The goal of our study was to determine if there was a measurable difference in the performance of experts and non-experts when it came to successfully completing the tasks that summary polar diagrams were created to address.
We decided to independently evaluate our dashboard, intentionally avoiding direct comparisons with other tools due to two reasons. On the one hand, existing implementations are primarily libraries rather than complete dashboards, without any of them integrating multiple visualization techniques as done in the proposed dashboard, and consequently, no real baseline currently exists for direct comparison.
Although some level of comparative analysis would, in principle, be possible using TRS, since it represents the only existing implementation that includes a limited degree of interactivity, this comparison is not practically feasible. TRS operates solely as a Python-based visualization library rather than a user-facing analytical environment, and therefore lacks an interactive interface for direct user engagement. Moreover, it provides no support for integrating multiple coordinated views or combining interactive elements such as overview+detail, aggregation, filtering, or Cartesian linking. As a result, TRS cannot reproduce or accommodate the complex interaction workflows and multi-idiom coordination that form the basis of our dashboard and the tasks evaluated in our user study.
On the other hand, our primary goal of the user study is to identify ways to improve the design and validate that the approach does indeed support both expert and non-expert users to accomplish their tasks. This approach aligns with the tasks and taxonomies outlined by Tory and Santos~\cite{user_study_argument_1, user_study_argument_2}, which further support our decision. By concentrating on a single tool, we could tailor the evaluation protocol to its unique features and capabilities, enabling a nuanced assessment of its strengths and weaknesses, as noted by Plaisant and Wu~\cite{user_study_argument_3, user_study_argument_4}.

Since the implementation of the small multiple technique on top of summary polar diagrams presented a slight extension of the idea presented in TRS, we did not evaluate its usefulness and effectiveness in a user study, as we did for the previously mentioned hybrid approach.

\subsection{Apparatus}
We conducted our study online by inviting participants to use and evaluate our proposed hybrid approach. The evaluation was enabled by two tools~---~the enhanced summary polar diagrams dashboard and the \textit{SoSci Survey}~\cite{sosci} online survey tool.
The dashboard allowed users to closely explore our implementation, interact with the data, and solve tasks. The SoSci Survey tool provided a platform to introduce participants to the study, collect sociodemographic data, train and test participants on multiple tasks that employed our hybrid approach, and evaluate its usability.

The study involved exploration tasks that required the utilization of all techniques employed in our combined approach. To properly showcase the designed dashboard layout, as presented in~\autoref{fig: Wine_comparison}, participants were instructed to ensure their screen resolution was at least $1800\times1080$ pixels. Additionally, we provided instructions on how to mitigate this limitation so that even participants with lower resolution screen sizes could participate.

\subsection{Participants}
The study involved a total of \totalParticipants participants after excluding those who did not complete more than 50\% of the provided questionnaire. Of these remaining participants, 11 identified as male, 9 identified as female, and 1 preferred not to declare their gender.
The participants' ages ranged from 20 to 50 years, with 62\% being between 30 to 40 years.
The majority, 66\%, held a Master's degree, with the Doctorate degree being the next most common at 28\%, and the remaining participants holding a Bachelor's degree. As for expertise, 52\% of participants selected one field, 42\% selected two, and 4\% (one participant) selected six different fields of expertise. The selections varied widely, with the most abundant being \textit{Sciences (Physical, Life, and Social)} at 40\%, followed by \textit{Information Technology} at 28\%, then \textit{Engineering and Technology} at 14\%, \textit{Healthcare and Medicine} at 8\%, and \textit{Business and Management}, \textit{Education and Teaching}, and \textit{Public Service and Administration} each at 3\%.
The sociodemographic structure, visualization expertise, and experience with polar coordinate charts are summarized in~\autoref{tab: participants}.
According to the subjective graph literacy test~\cite{graph_literacy_test}, participants' mean score was 24.76 (S.D. = 4.88) out of 30. The average time taken to complete both the train and test tasks was 20 minutes (S.D. = 5 minutes), and is discussed further in~\autoref{subsec: study_results}.
We also asked about their prior experience with polar charts and summary polar diagrams specifically. The results indicate that 71\% percent had never used polar charts, and 95\% percent had never utilized summary polar diagrams.

\begin{table}
    \centering
    {\rowcolors{2}{}{gray!15}
    \begin{tabular}{lcclccc}
        \toprule
         \# & Gender & Age & Degree & Years of Exp. & P.C. Exp. & S.P.D. Exp. \\
        \midrule
        P0 & \faMale & 20 -- 30 & Master’s & \textless~5 & No & No \\
        P1 & \textbf{NA} & 30 -- 40 & Doctorate & 5 -- 10 & No & No \\
        P2 & \faFemale & 30 -- 40 & Doctorate & 5 -- 10 & Yes & No \\
        P3 & \faMale & 30 -- 40 & Doctorate & 5 -- 10 & No & No \\
        P4 & \faMale & 30 -- 40 & Master’s & \textless~5 & No & No \\
        P5 & \faMale & 20 -- 30 & Master’s & \textless~5 & No & No \\
        P6 & \faFemale & 40 -- 50 & Doctorate & 20 -- 25 & Yes & No \\
        P7 & \faFemale & 30 -- 40 & Master’s & 10 -- 15 & No & Yes \\
        P8 & \faMale & 30 -- 40 & Master’s & 5 -- 10 & No & No \\
        P9 & \faFemale & 30 -- 40 & Master’s & 5 -- 10 & No & No \\
        P10 & \faMale & 30 -- 40 & Master’s & 5 -- 10 & Yes & No \\
        P11 & \faMale & 30 -- 40 & Master’s & 10 -- 15 & No & No \\
        P12 & \faFemale & 20 -- 30 & Master’s & 5 -- 10 & No & No \\
        P13 & \faFemale & 30 -- 40 & Doctorate & 5 -- 10 & Yes & No \\
        P14 & \faMale & 30 -- 40 & Master’s & 5 -- 10 & No & No \\
        P15 & \faMale & 20 -- 30 & Master’s & \textless~5 & No & No \\
        P16 & \faMale & 30 -- 40 & Master’s & 5 -- 10 & Yes & No \\
        P17 & \faFemale & 30 -- 40 & Master’s & \textless~5 & No & No \\
        P18 & \faFemale & 40 -- 50 & Doctorate & 15 -- 20 & Yes & No \\
        P19 & \faMale & 20 -- 30 & Master’s & \textless~5 & No & No \\
        P20 & \faFemale & 20 -- 30 & Bachelor’s & \textless~5 & No & No \\
        \bottomrule
    \end{tabular}}
    \caption{\textit{\textbf{Participant demographics}. Exp., P.C., and S.P.D. stand for experience, polar charts, and summary polar diagrams, respectively. In addition to the data presented in the table, we also asked participants about their expertise and to subjectively evaluate their graph literacy.}}
    \label{tab: participants}
\end{table}

\subsection{Experimental Factors}
\label{subsec: experimental factors}
Our study involved two types of experimental factors~---~task-related~(TR) and user-centric~(UC) factors. We describe each of them below.

\begin{itemize}
    \item \textbf{[TR] Types of analytical tasks}: Participants were asked to complete eight different analytical tasks that varied in complexity. These tasks were designed to reflect real-world scenarios commonly encountered when using summary polar diagrams. Our goal was to thoroughly cover the range of analytical activities one might perform with these diagrams, thereby exhausting the space of possible tasks that could be solved using the proposed hybrid approach.
    \item \textbf{[TR] Complexity levels of questions asked}: Each task in the sequence increased in complexity, with the first task being the simplest and the last task being the most complex. All eight tasks were split into four pairs, with each pair addressing different analytical and interaction requirements.
    \item \textbf{[TR] Interaction Requirements}: To assess the usability of all techniques present in the hybrid approach, each task required different levels and types of interaction with the dashboard. While some tasks required filtering via the legend, inspecting the detail diagram using the tooltip, or interacting with the Cartesian-Linking plot, others required semantic zooming by brushing over the overview diagram.
    \item \textbf{[UC] Level of guidance or instructions provided}: As stated earlier, the tasks were organized to address analytical problems of varying complexity, from the easiest to the most complex. The tasks within each pair followed the same logical progression, starting with an easier, train-oriented task that provided feedback, and culminating in a more challenging, testing-oriented task that did not include any feedback. The feedback in the train task informed participants whether their answers were correct or not. This structure allowed us to assess participants' abilities as they moved from simpler to more complex analytical scenarios.
\end{itemize}

\subsection{Experimental Design}
\label{subsec: experimental design}
The order of pages in the questionnaire and the tasks within each page were fixed for all participants P0~--~P20. The rationale behind this decision was to ensure that each participant was subjected to the same training procedure, regardless of their visualization expertise or experience with summary polar diagrams. However, in order to reduce \textit{order bias} and prevent \textit{straight-lining}, the six answer choices for each task in the questionnaire were randomized.

Before each task pair, the questionnaire presented a tutorial page that explained the important elements required for solving the upcoming tasks. Each tutorial page consisted of multiple annotated screenshots, each isolating a specific element of the dashboard or diagrams. These screenshots were accompanied by concise descriptions explaining their significance. We strived to make each tutorial page as informative yet as lean as possible to mitigate potential \textit{participant fatigue} and keep participants prepared for the tasks ahead. Tutorial screenshots incorporated diagrams with the climate data set, as described in~\autoref{subsec: climate}, while the tasks involved the wine data set, as described in~\autoref{subsec: wine}. We deemed this separation necessary to mitigate potential \textit{learning (practice) effect}. Moreover, to mitigate \textit{expectation bias}, we structured our train-test pairs to address the same analytical and interaction requirements while being completely different in content.

\subsection{Metrics and Analysis}
We collected the following data for each participant:
\begin{itemize}
    \item \textbf{Sociodemographic}: Gender, age group, highest degree, field of experience, and years of experience.
    \item \textbf{Visualization-relevant}: Subjective graph literacy test, experience with charts in polar coordinate system, and experience with summary polar diagrams such as the Taylor or mutual information diagrams.
    \item \textbf{System-relevant}: System usability scale~(SUS)~\cite{sus-scale}, time to completion, and optional feedback.
\end{itemize}

\subsection{Tasks}
\label{subsec: tasks}
Our experiment involved eight tasks, split into four pairs, with each pair addressing a different analytical problem. We designed our tasks so that each subsequent pair required new interactive actions or employed parts of the dashboard not used earlier. Tasks within each pair maintained consistent dashboard elements and interactive actions but increased in complexity, starting with a simpler training task and ending with a more complex test task. The subsequent part of this section will describe the four distinct task pairs and their elements, presenting them in order from easiest to most complex, while also connecting them to the experimental factors presented in~\autoref{subsec: experimental factors}.

\begin{itemize}
    \item \textbf{Model similarity analysis}: Participants identified the model closest to, and the three models furthest from, a reference model. This task utilized summary polar diagrams, requiring only tooltip interaction for precise value examination.
    \item \textbf{Cluster identification and quantification}: Participants determined the total number of clusters and the model count within a specific cluster. This task necessitated the use of the overview diagram and/or main legend.
    \item \textbf{Intra- and inter-cluster examination}: In the first task, participants selected a specific cluster and identified its central model based on variation of information. The second task involved selecting an additional cluster and determining which model from the first cluster was most distant from the second. These tasks required either brushing the overview diagram or employing multi-selection in the main legend.
    \item \textbf{Advanced information theoretic analysis}: This most complex pair began with participants selecting two models and calculating their combined entropy sum. The second task involved choosing two clusters and determining their variation of information ranges, from lowest to highest. These tasks demanded legend multi-selection and utilization of the Cartesian-linking plot.
\end{itemize}

\subsection{Procedure}
All participants P0~--~P20 received an invitation e-mail with a link to the questionnaire, briefly explaining the user study's purpose, duration, and emphasizing that no prior experience with polar coordinate charts or summary polar diagrams was necessary.
The questionnaire consisted of four distinct parts.
First, upon clicking the questionnaire link, users encountered an introductory page detailing the study, data collection practices, and their right to exit at any time. This part also included the collection of sociodemographic and visualization-relevant data.
Second, participants were invited to open a dashboard in a new browser tab and proceed only after confirming it matched the provided image.
Third, a cycle of four tutorial, training task, and test task pages followed, as described in~\autoref{subsec: experimental design}. For each set, participants \textbf{a)} examined a tutorial page preparing them for upcoming tasks, \textbf{b)} solved a training task with feedback on their answer, allowing error correction without consequences, and \textbf{c)} solved a test task without any feedback.
Fourth, after completing the cycle, participants encountered a final page containing the system usability scale question and an optional feedback text field.

\subsection{Results}
\label{subsec: study_results}

This section contains an overview of both quantitative and qualitative results. All analyses were conducted using Python (version 3.8.10).

\subsubsection{Quantitative Results}
As noted earlier in the text, participants took an average of 20 minutes to complete the combined train and test tasks. For almost all train-test pairs, users required around 10\% more time to finish the test task than the train task. The only exception is with the third task, where users spent on average 50\% more time during train than the test. As this pair is about exploring the interactive abilities of the system, we believe the users used their time liberally during train to try out all the possibilities of the system, which they then utilized to quickly finish the upcoming test task.

\begin{figure}[htb]
    \centering
    \begin{subfigure}[t!]{0.4\linewidth}
        \centering
        \includegraphics[width=\linewidth]{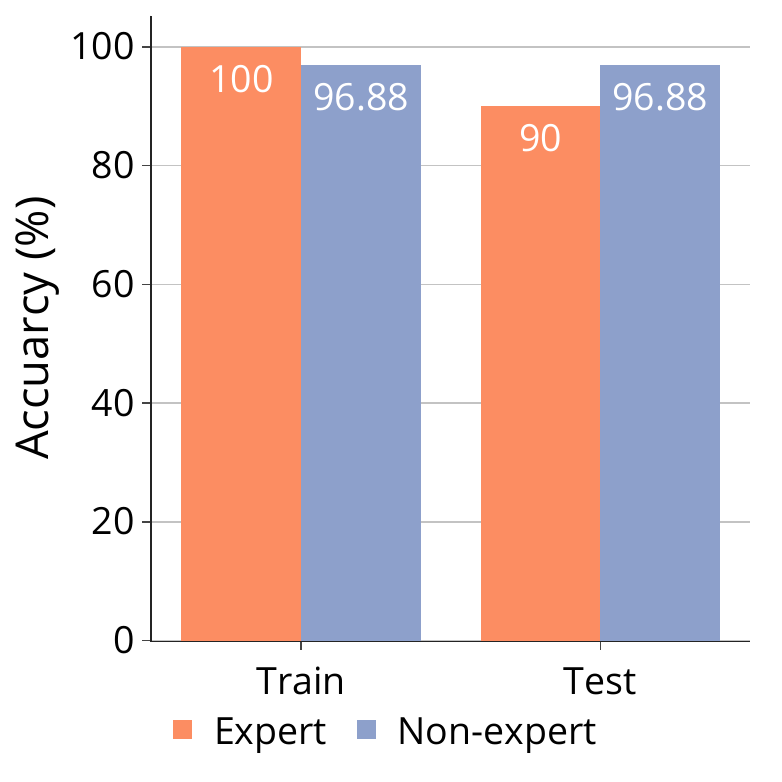}
        \label{fig: Study_accuracy}
    \end{subfigure}
    \begin{subfigure}[t!]{0.4\linewidth}
        \centering
        \includegraphics[width=\linewidth]{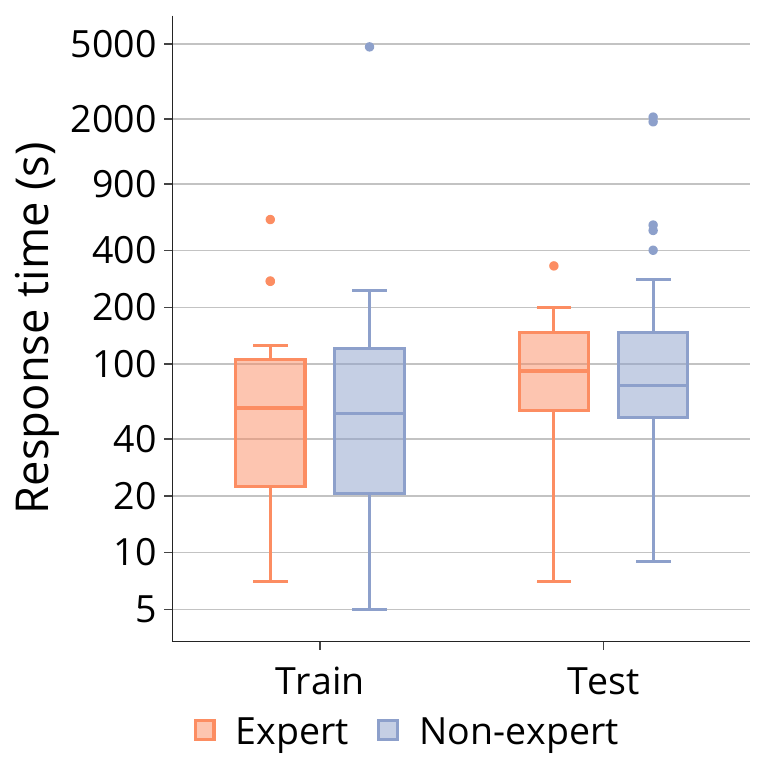}
        \label{fig: Study_response_times}
    \end{subfigure}
    \caption{\textit{\textbf{Accuracy and response time for experts and non-experts}. The y-axis of the right figure is presented on a logarithmic scale to better visualize the wide range of values. The results indicate that the experts achieved perfect accuracy on the train tasks, with all responses classified as correct. However, their performance declined on the test tasks, where they produced a mix of correct and incorrect responses. In contrast, the non-expert group demonstrated consistent accuracy for both train and test tasks, resulting in a uniform classification of correct or incorrect.  In contrast, the response times were comparable between the groups.}}
    \label{fig: Study_results}
\end{figure}

To test our main hypothesis, we categorized participants into two groups~---~expert or non-expert~---~based on their subjective graph literacy scores. We used the third quartile of the scores as the cutoff value, with participants scoring below this considered non-experts and those scoring above it considered experts.
The results presented in~\autoref{fig: Study_results} indicate no significant difference in accuracy or response times between the expert and non-expert participant groups.
A comprehensive statistical analysis was conducted to compare the performance of experts and non-experts in both train and test conditions. Prior to the main analysis, the normality of the data distribution was assessed using both Shapiro-Wilk and Kolmogorov-Smirnov tests.
For the train data, the Shapiro-Wilk test for experts yielded $W = 0.653$ with $p = 1.098\times 10^{-5}$, and the Kolmogorov-Smirnov test showed $D = 1.0$ with $p = 2.782\times 10^{-238}$. For non-experts, the Shapiro-Wilk test resulted in $W = 0.1689$ with $p = 3.236\times 10^{-17}$, and the Kolmogorov-Smirnov test showed $D = 1.0$ with $p = 0.0$.
For the test data, the Shapiro-Wilk test for experts yielded $W = 0.8975$ with $p = 0.0370$, and the Kolmogorov-Smirnov test showed $D = 1.0$ with $p = 2.782\times 10^{-238}$. For non-experts, the Shapiro-Wilk test resulted in $W = 0.373$ with $p = 4.712\times 10^{-15}$, and the Kolmogorov-Smirnov test showed $D = 1.0$ with $p = 0.0$.
These results consistently indicated significant deviations from normality for both groups in both conditions (all $p < 0.05$). Given these violations of normality assumptions, non-parametric Mann-Whitney U tests were employed for the main analyses.

For the training condition, the Mann-Whitney U test revealed no significant difference between experts and non-experts ($U = 639.5$, $p = 1.0$). Experts averaged $101.0$ seconds, while non-experts averaged $147.61$ seconds (with pooled $SD = 527.05$). The effect size, calculated using the rank-biserial correlation, was negligible ($r = 0.0008$).
Similarly, in the testing condition, no significant difference was found between the two groups ($U = 651.0$, $p = 0.9122$). Experts averaged $104.15$ seconds, compared to non-experts at $169.28$ seconds (with pooled $SD = 304.77$). Again, the effect size was negligible ($r = -0.0172$).
These results consistently indicate no statistically significant differences in performance between experts and non-experts across both training and testing conditions. The negligible effect sizes further support the lack of practical significance in the observed differences. It's worth noting that while the mean times for non-experts were consistently higher than those for experts ($46.61$ seconds longer in train and $65.13$ seconds longer in test), these differences did not reach statistical significance given the high variability in the data and the non-normal distributions.
The use of non-parametric tests and appropriate effect size measures ensures the robustness of these findings, accounting for the non-normal distribution of the data. However, it's important to consider that the lack of significant differences could be due to factors such as small sample sizes or high within-group variability, rather than a true absence of difference between expert and non-expert performance.

\begin{figure*}[tb]
    \centering
    \includegraphics[width=0.8\linewidth]{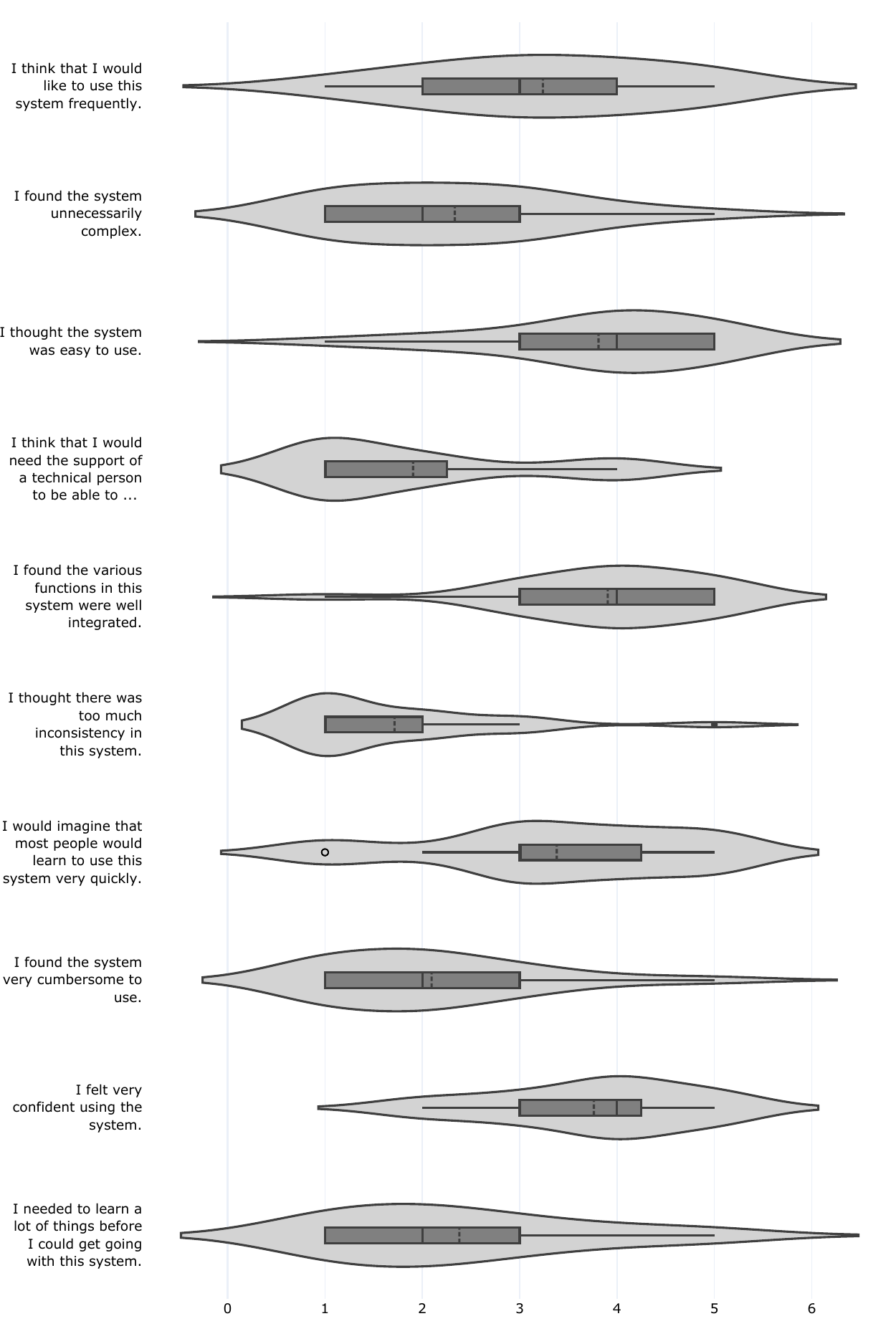}
    \caption{\textit{\textbf{System Usability Scale (SUS) scores from user study participants}. We can see that the majority of participants evaluated the system as highly usable and well integrated. However, the results also show disagreement among users regarding whether they would learn to use this system quickly, suggesting there are still opportunities to further improve certain aspects of the system's design and functionality. The mean SUS score is 69.16, and the median 75.}}
    \label{fig: SUS_results}
\end{figure*}

\begin{figure*}[tb]
    \centering
    \begin{subfigure}[t!]{0.45\linewidth}
        \centering
        \includegraphics[width=\linewidth]{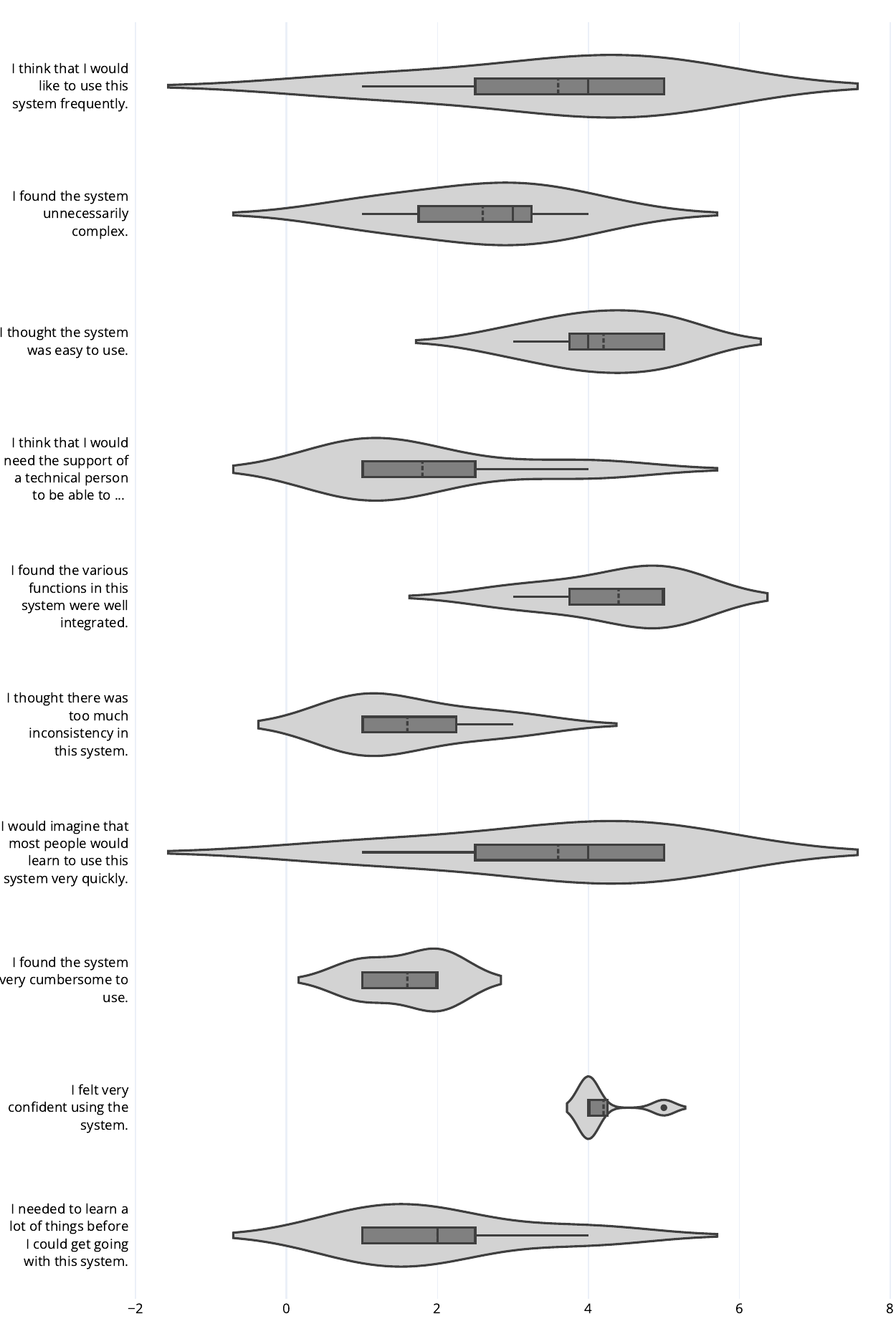}
    \end{subfigure}
    \hfill
    \begin{subfigure}[t!]{0.45\linewidth}
        \centering
        \includegraphics[width=\linewidth]{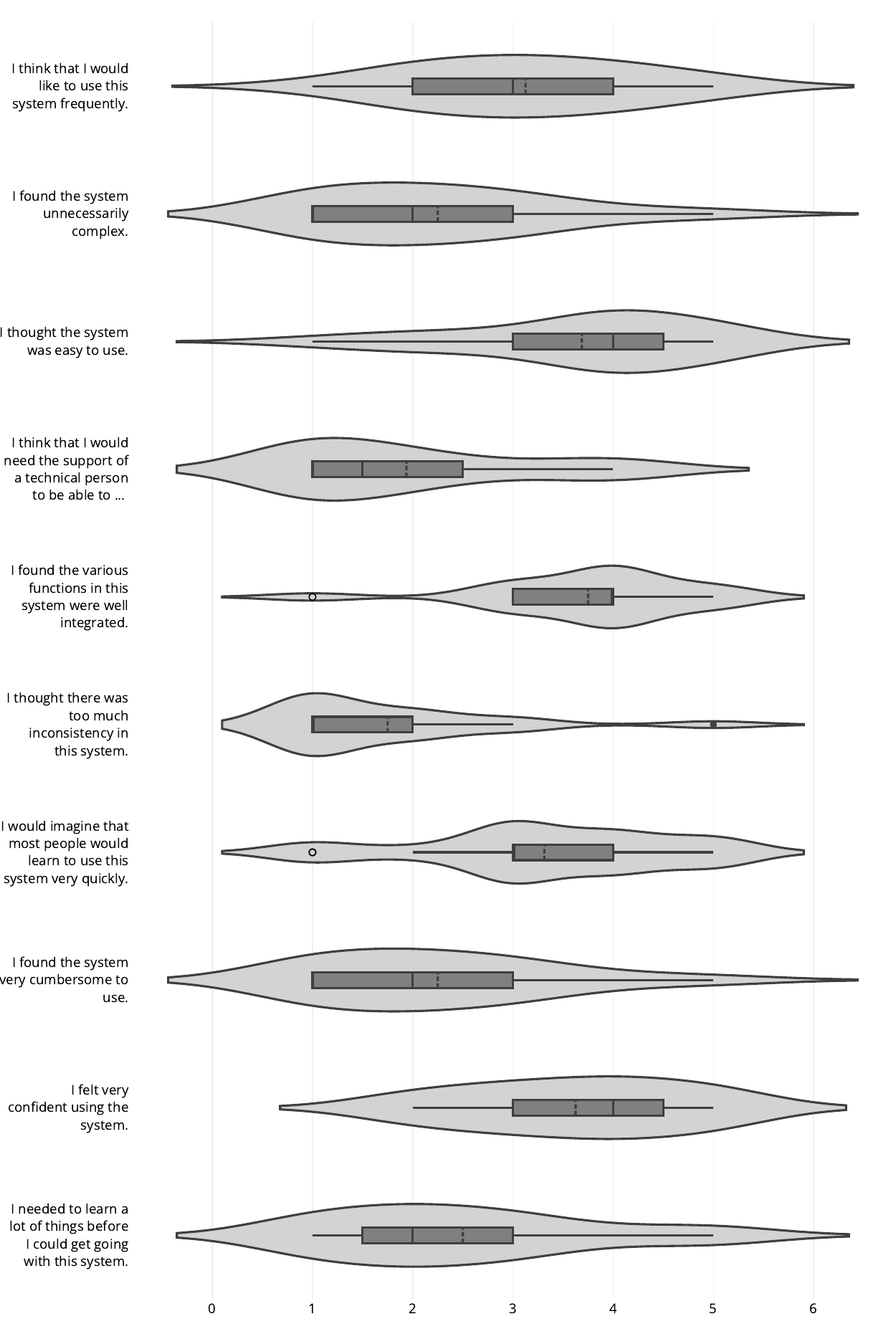}
    \end{subfigure}
    \caption{\textit{\textbf{Stratified System Usability Scale (SUS) scores from user study participants.} Results are stratified by user expertise with experts on the left (mean SUS score 76, median 77.5) and non-experts on the right (mean SUS score 67.03, median 70).}}
    \label{fig: SUS_results_stratified}
\end{figure*}

The last page of our questionnaire contained the System Usability Scale (SUS), where users answered ten different questions on a scale of 1 to 5. The overall trend indicates that the system is well-integrated, easy to use, and without too many inconsistencies. The only notable disagreement among users can be seen with the question ``I would imagine that most people would learn to use this system very quickly'', where we observed scores ranging across both ends of the scale. The mean SUS score is 69.16, and the median 75. All results can also be seen in~\autoref{fig: SUS_results}.
Additional analyses of the SUS results, stratified by user expertise (expert \textit{vs.} non-expert), demonstrated both similarities and differences between the two groups. The expert group's responses contained less variability. For example, the answers to the statements ``I found the system very cumbersome to use'' and ``I felt very confident using the system'' were more heterogeneous in the non-expert group than the expert group. However, the overall mean and median score values across the two user groups were largely similar but slightly higher for the experts (expert mean SUS score 76.0, median 77.5; non-experts mean SUS score 67.03, median 70), as observed in~\autoref{fig: SUS_results_stratified}.

\subsubsection{Qualitative Results}
Our questionnaire also included an optional text field where users could provide feedback about the system, the study, or any inconsistencies they encountered while participating. Here we summarize all the experiences and open-ended feedback shared by the participants.

\paragraph{Data Visualizations.} We observed an agreement between participants regarding the value of the integrated techniques. P0 noted that ``\textit{I think the idea is very good and I [prefer] comparing the proximity via the radial scale [... which] is easier than looking at barplots.}''. P4 and P18 supported the easy comprehension of the diagrams by stating ``\textit{[...] everything looks well-developed and quite straightforward to follow}'', and ``\textit{Extremely user-friendly questionnaire with a visually appealing design!}'', respectively. Furthermore, some participants also remarked on the overview diagram and its connection with the main legend. Notably, P10 remarked ``\textit{The order of clusters (3, 2, 4, 1, 5, 6) was confusing}'', while P0 said ``\textit{[...] I think it's also a little confusing that points from different clusters have different colors [...]}''. We observed suggestions from multiple participants regarding the overview diagram and the aggregation technique through clustering, thus proving that this combination can be further improved. 

\paragraph{Interactive Features.} Even though we provided instructions on interactive actions, some participants still encountered problems during certain tasks. P0 remarked ``\textit{I think the part with the clustering and comparing the clusters could be improved. It felt a little clumsy and it took some time to select the clusters I was interested in.}''. P7 and P12 echoed this sentiment with ``\textit{I missed a button or clear instructions on resetting the page and returning to the initial state.}'' and ``\textit{Double-clicking on a data point to select/de-select all other models of the corresponding cluster didn't work in my case [...]}'', respectively. These participant remarks regarding the inherent interactive behaviors of the Plotly library suggest opportunities to enhance the intuitiveness of these interactions or introduce clearer guidance on how to utilize them effectively.

\paragraph{System Usability.} Only one participant reported an issue that could be classified as a system-level problem. Notably, P8 said ``\textit{I had an issue where some of the models would remain in the bottom left 1D diagrams when I deselect them too quickly}''. Our internal testing indicated that this problem appeared in specific cases when the user accessed our dashboard online using a lower-end system, rather than running it locally. In these situations, a problem could occur when user interactions did not reliably trigger the corresponding events. However, our testing also showed that this problem never arose when the dashboard was run locally, suggesting the issue was likely due to latency or performance limitations of the online deployment.

\section{Application Examples}
\label{sec: application examples}

The implemented dashboard integrates the proposed techniques with summary polar diagrams across three different domains: climate, data science, and machine learning~(ML). The climate domain is the one for which the original Taylor diagram was initially developed and in which it is extensively used. The use case from the data science domain has not been explored before in the context of summary polar diagrams, while the machine learning use case was initially explored in TRS, but not with the small multiple technique proposed in this work.

\subsection{Climate Model Comparison}
\label{subsec: climate}

To demonstrate the capabilities of the newly developed diagram, \textcite{taylor-diagram-original-paper} conducted a comparison of various climate models and their individual features, including sea level pressure, surface air temperature, and total cloud fraction, among others. Comparing statistical and information-theoretic measures of climate models is important in order to determine which model overemphasizes or underestimates the magnitudes of the mentioned features, which model's cycles align with the observed data (showing good correlation or mutual information), and which models perform better or worse overall.
We applied our ideas to this problem as shown in Supplemental Video 1 and \autoref{fig: Climate}. In our example, we focused on air temperature using observational (reference) data from \textcite{climate-observation} and simulated model data from the Coupled Model Intercomparison Project Phase~5~(CMIP5)~\cite{climate-cmip5}.
Our hybrid approach allows us to detect two distinct clusters, with the second one containing models that make the worst air temperature predictions. Moreover, after inspecting the standard deviations, we could see that all models grossly underestimate the amplitude of the air temperature cycle. 
However, even though being one of the worst models according to the CRMSE, the model named \textit{CanESM2} is the one making the least error. Furthermore, if we inspect the correlation, we can see that models \textit{GISS-E2-R} and \textit{CNRM-CM5}, besides being the best, also have the best phasing of the annual air temperature cycle.

\begin{figure*}[htb]
    \centering
    \begin{subfigure}[t!]{0.2\textwidth}
        \centering
        \includegraphics[width=\textwidth]{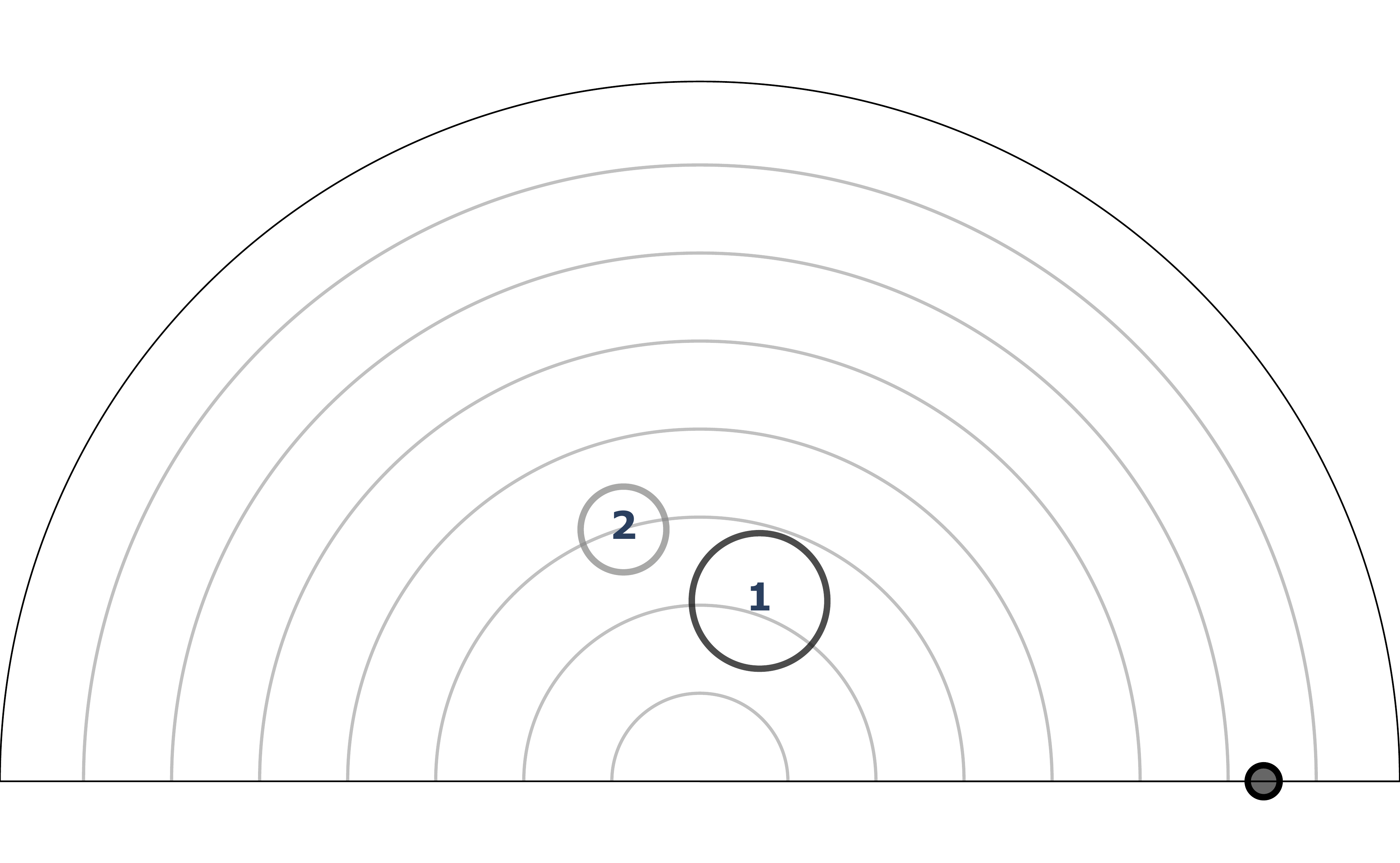}
        \includegraphics[width=\textwidth]{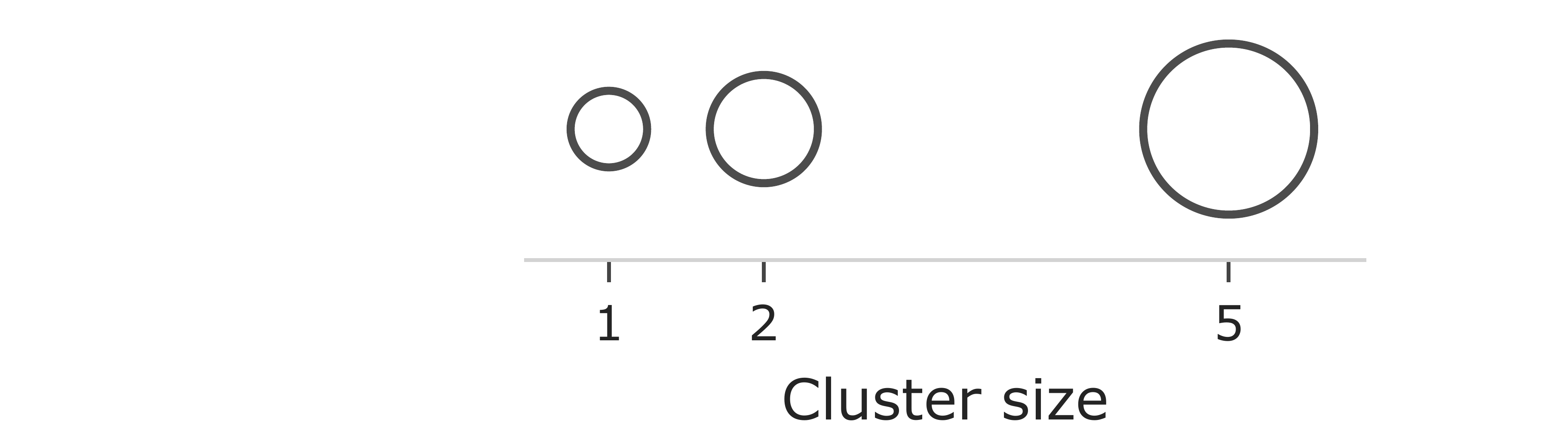}
        \includegraphics[width=\textwidth]{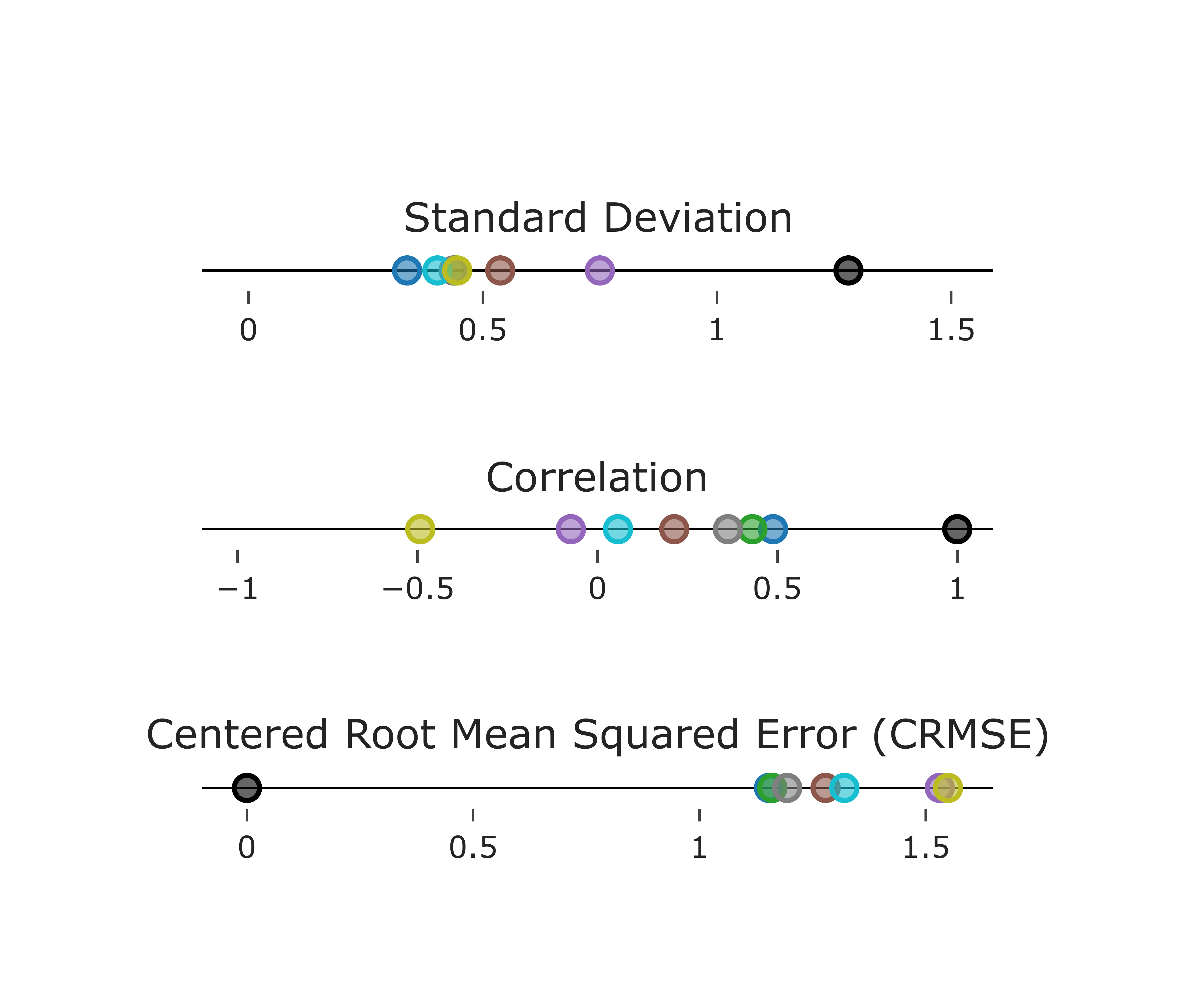}
        \label{fig: Climate_overview}
    \end{subfigure}
    \hfill
    \begin{subfigure}[t!]{0.7\textwidth}
        \centering
        \includegraphics[width=\textwidth]{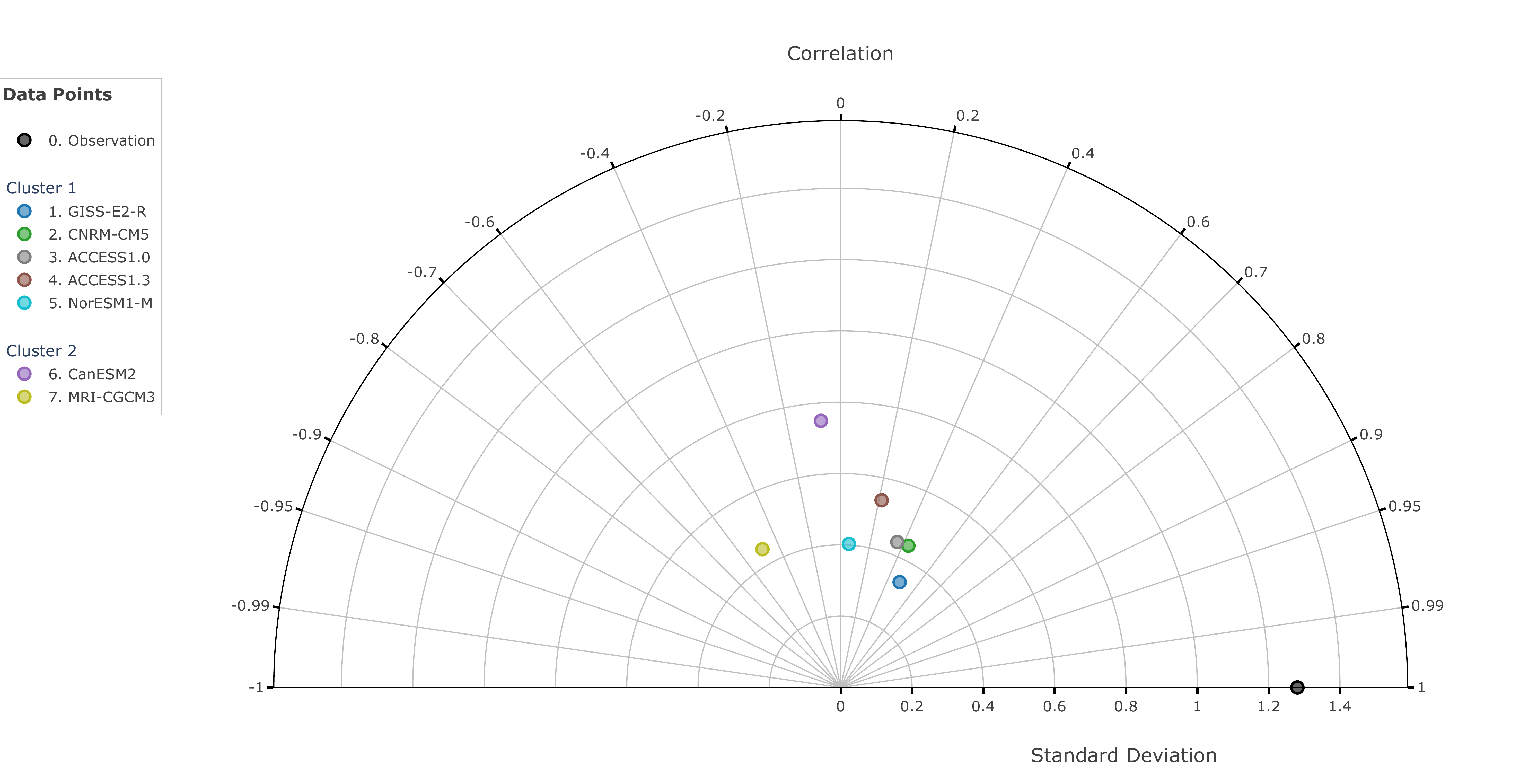}
        \label{fig: Climate_detail}
    \end{subfigure}
    \caption{\textit{\textbf{Enhanced Taylor diagrams using overview+detail, aggregation, Cartesian linking, and interactive filtering.} The initial view of the enhanced diagrams visualizes the air temperature predictions of seven climate models compared to the observed values before any interactive actions are taken. These example diagrams were used for train tasks in the user study.}}
    \label{fig: Climate}
\end{figure*}

\subsection{Sample Analysis}
\label{subsec: wine}

Sampling is a critical aspect of research, as it directly impacts the efficiency, accuracy, and validity of the findings. Analyzing the entire population can be time-consuming and costly, which is why researchers often rely on sampling techniques to save time and resources~\cite{sampling-2}. However, the sampling process can significantly influence the scientific rigor and practical impact of the research~\cite{sampling-1}. The example shown in~\autoref{fig: Wine_comparison} and given as a test to our user study participants, utilizes the wine data set by \textcite{wine-dataset}.
The data set consists of numerous wine samples, each including eleven continuous and one categorical feature. In order to avoid analyzing the entire data set, we decided to perform stratified sampling and analyze the selected wine samples against the theoretical wine sample containing median property values. This procedure created nineteen wine samples plus the median, while our enhanced implementation clustered them into six distinct clusters according to their information-theoretic properties. 
The hybrid approach here allowed us to closely examine each wine representative, compare their intra- and inter-cluster differences, and analyze if these clusters had any correlation with the quality of each wine (the categorical property of the data set). Our analysis shows that while some wine samples are grouped by quality, anomalies exist requiring further examination.

\subsection{Machine Learning Hyper-parameter Tuning}

\begin{figure*}[htb]
    \begin{subfigure}[t]{0.3\linewidth}
        \centering
        \includegraphics[width=\linewidth]{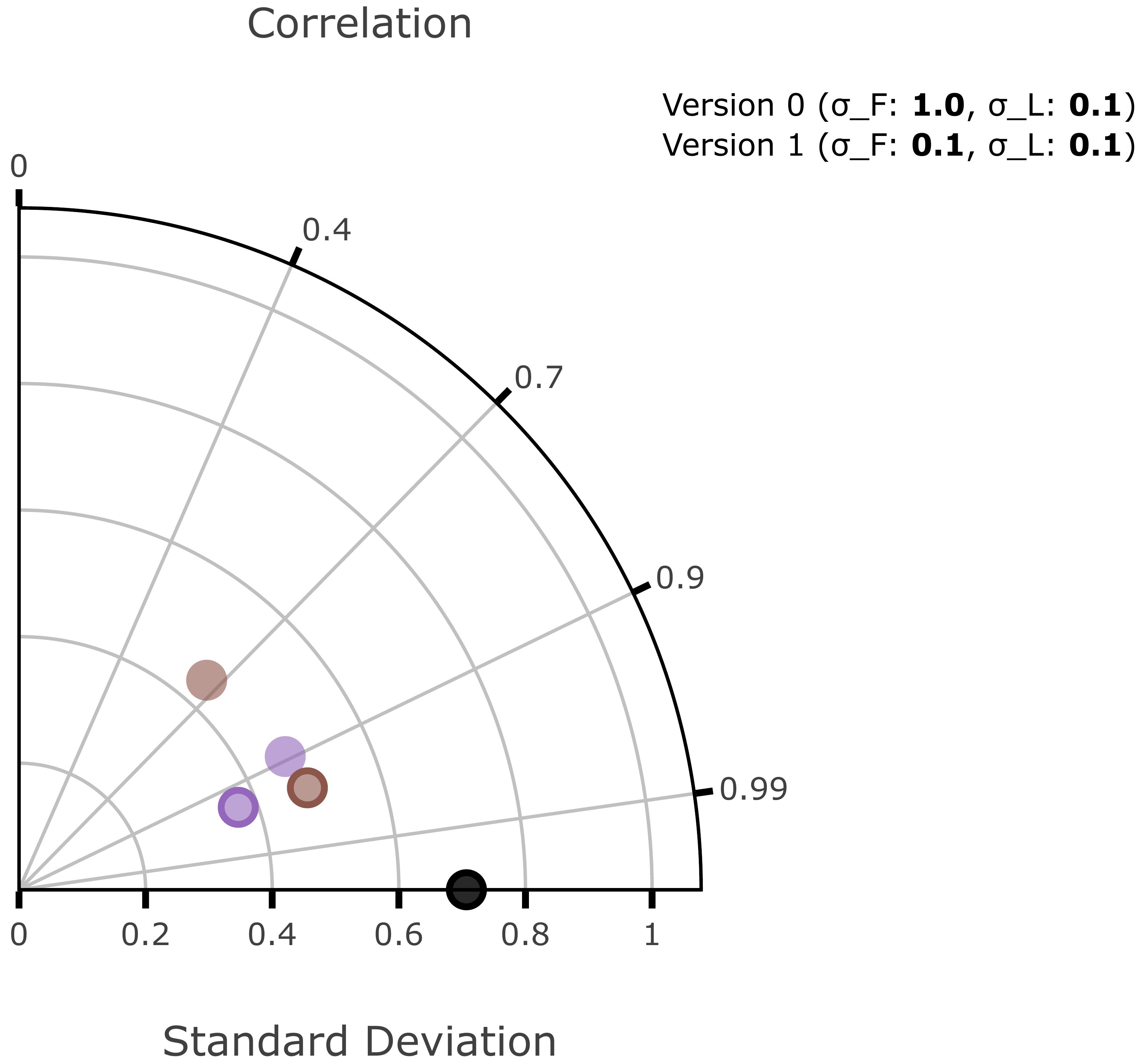}
    \end{subfigure}
    \hfill
    \begin{subfigure}[t]{0.3\linewidth}
        \includegraphics[width=\linewidth]{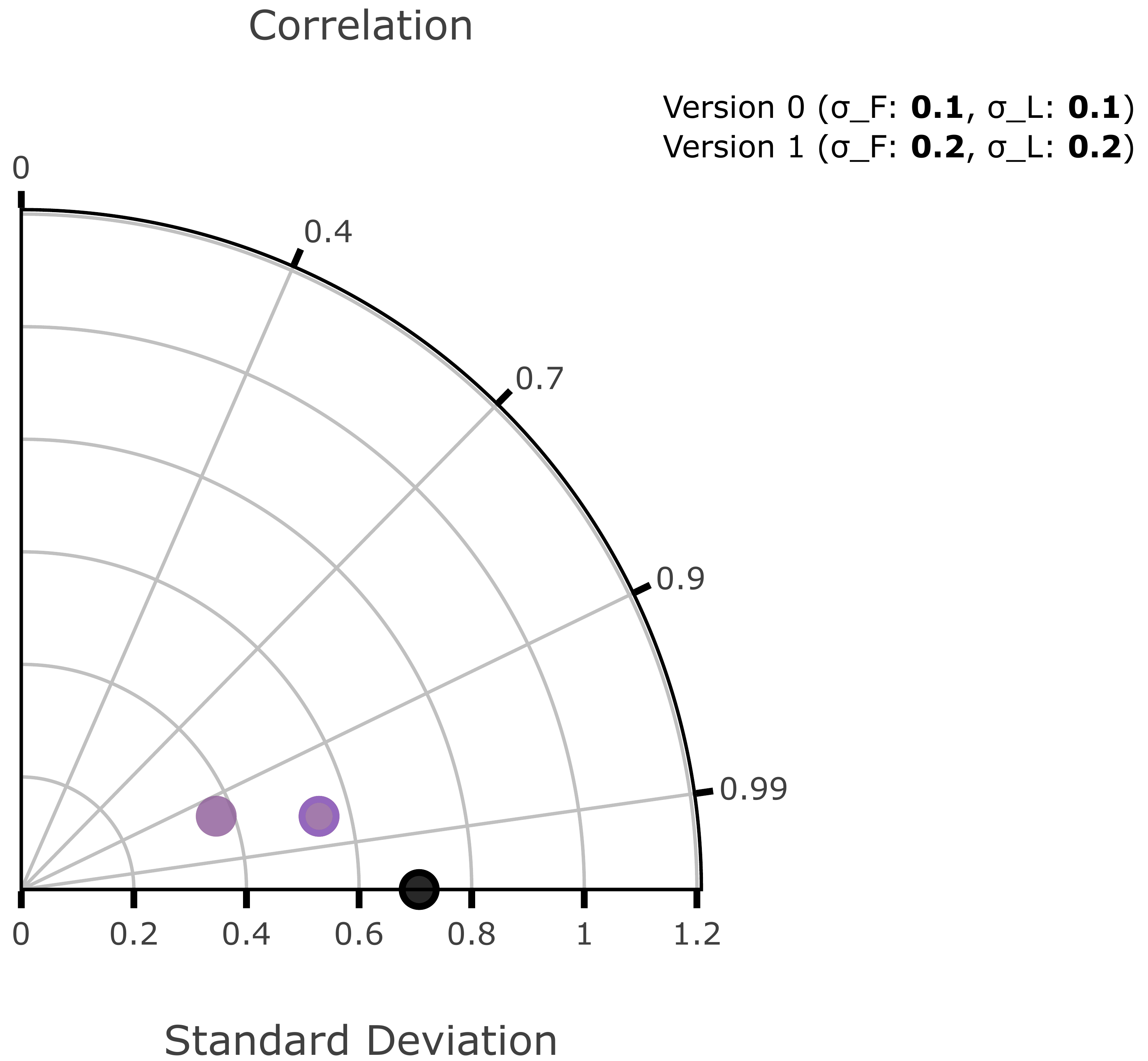}
    \end{subfigure}
    \hfill
    \begin{subfigure}[t]{0.3\linewidth}
        \includegraphics[width=\linewidth]{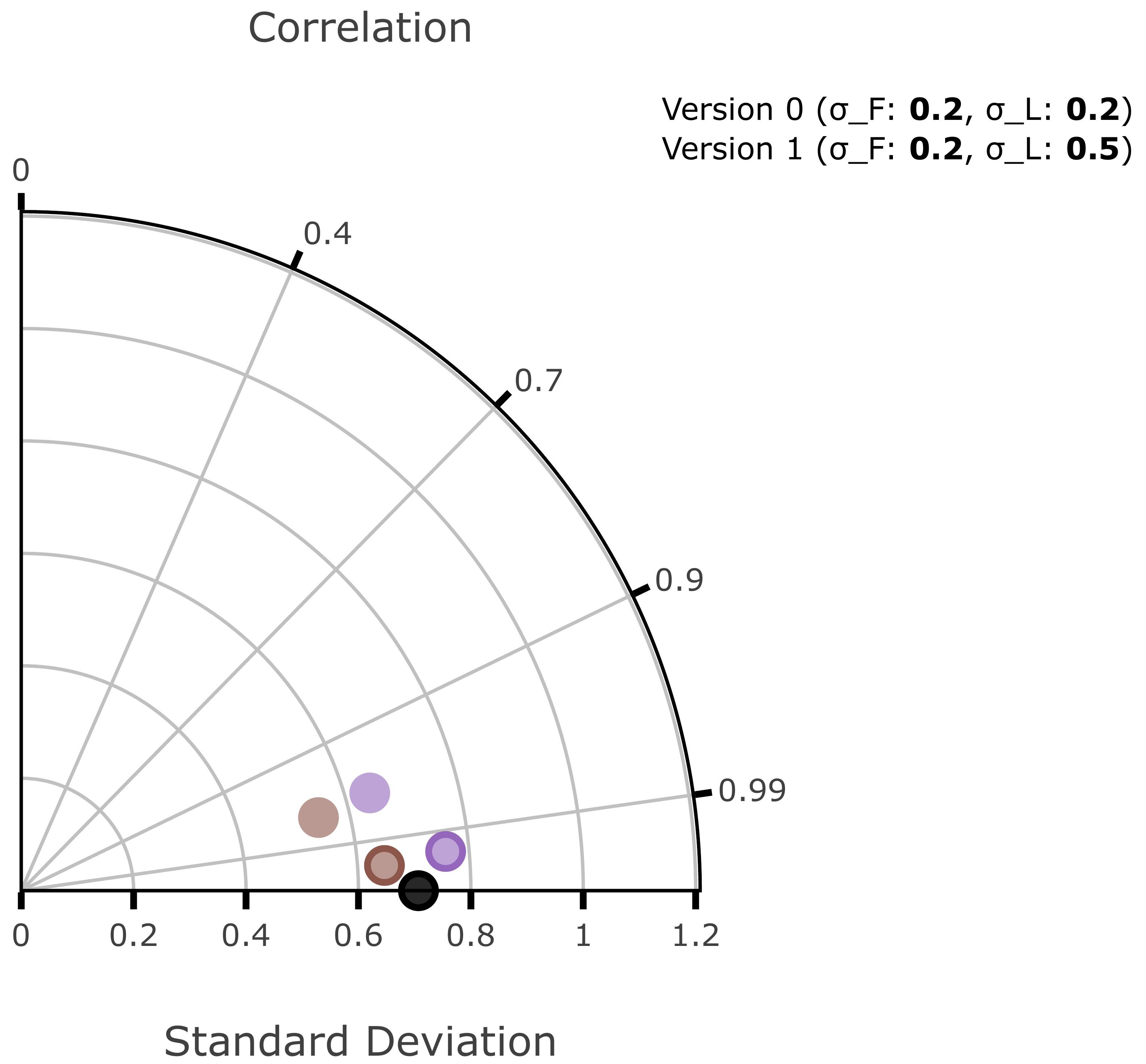}
    \end{subfigure}

    \bigskip
    \bigskip
    
    \begin{subfigure}[t]{0.3\linewidth}
        \includegraphics[width=\linewidth]{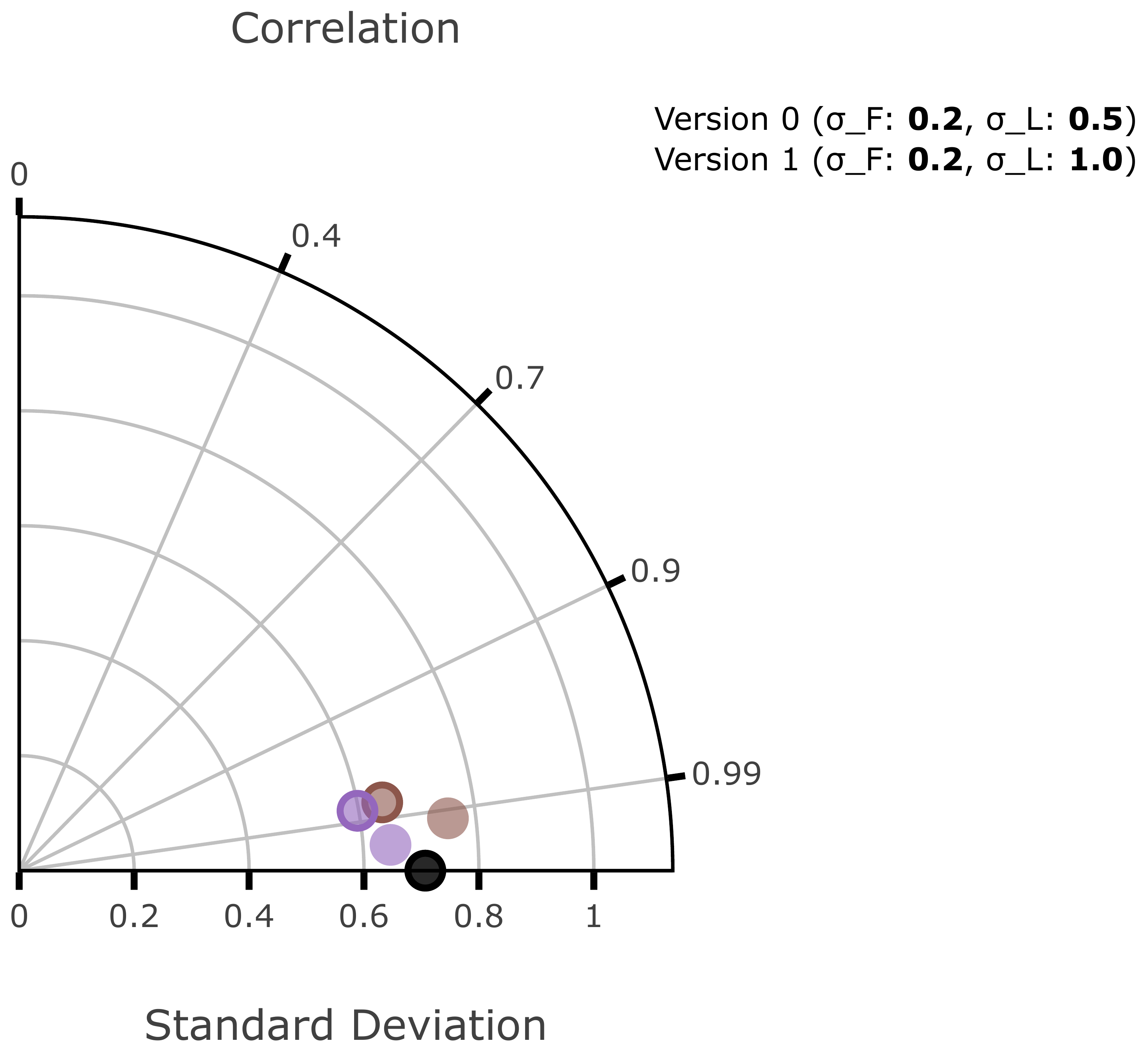}
    \end{subfigure}
    \hfill
    \begin{subfigure}[t]{0.3\linewidth}
        \includegraphics[width=\linewidth]{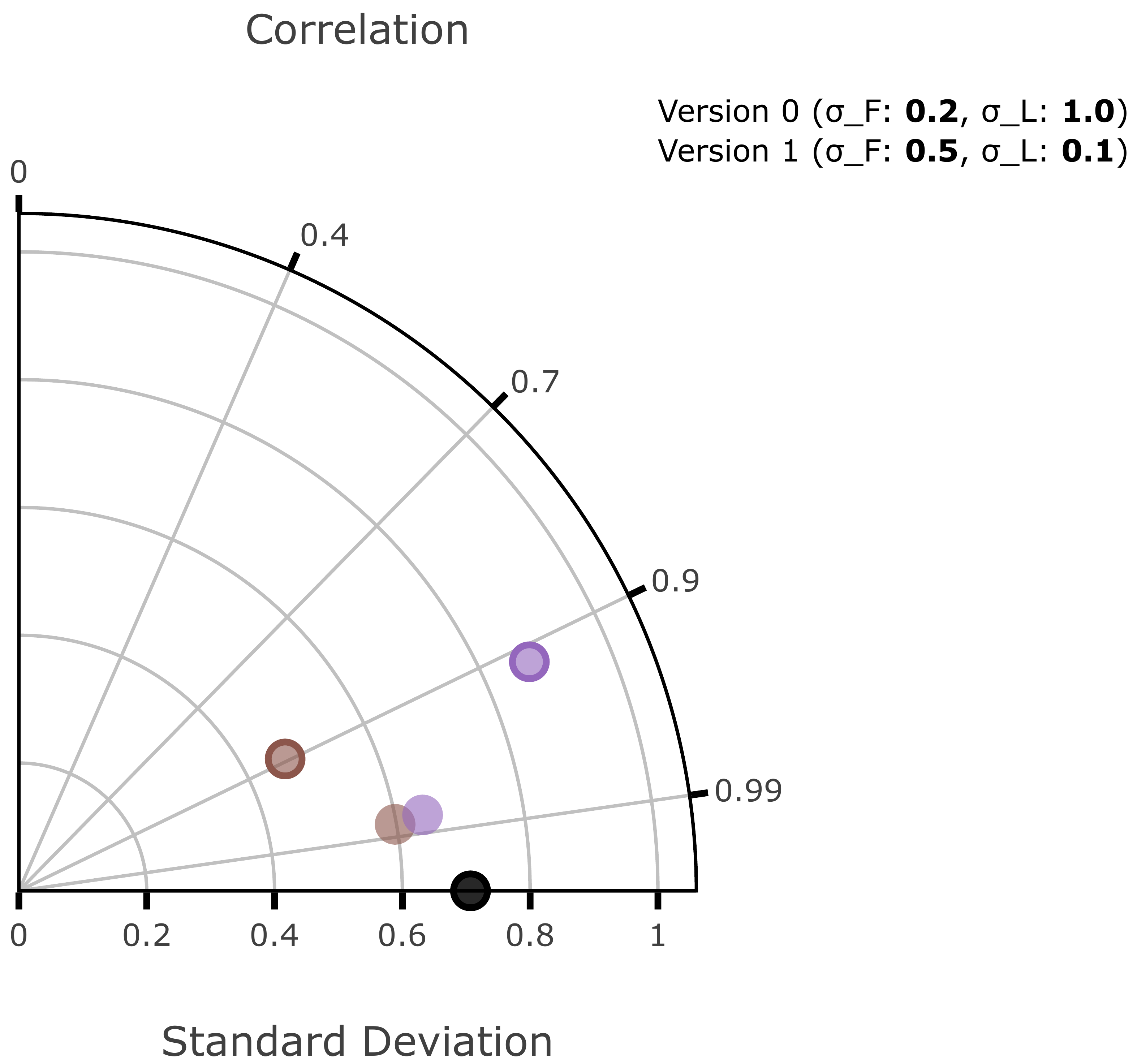}
    \end{subfigure}
    \hfill
    \begin{subfigure}[t]{0.3\linewidth}
        \includegraphics[width=\linewidth]{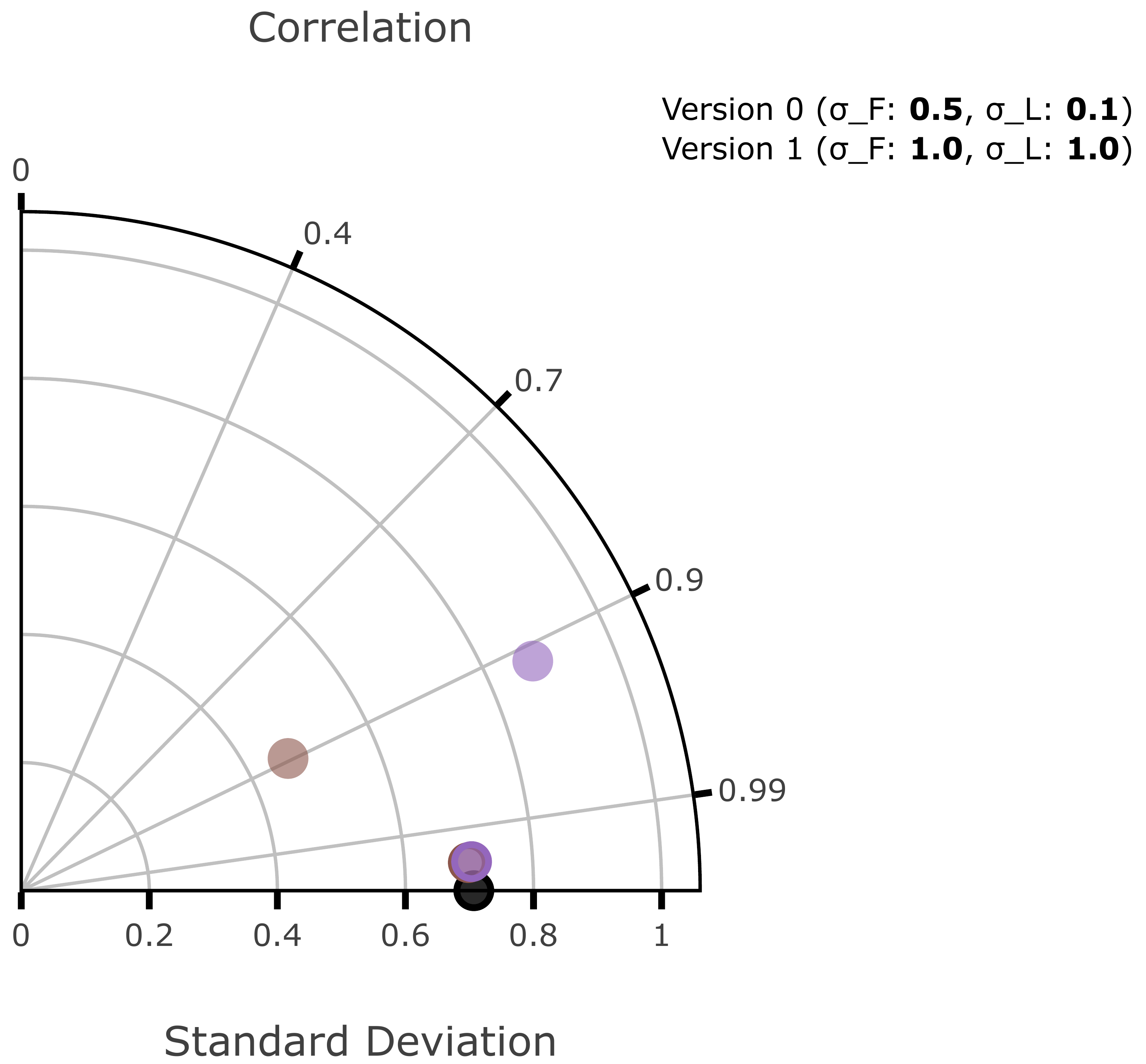}
    \end{subfigure}
    
    \begin{subfigure}[l]{0.3\linewidth}
        \includegraphics[width=0.9\linewidth]{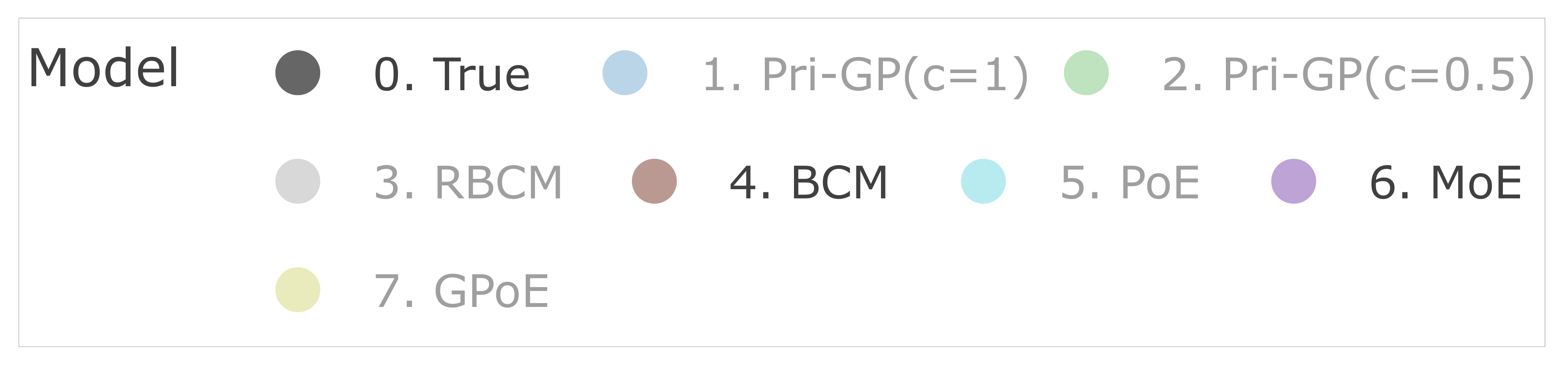}
    \end{subfigure}\hfill

    \caption{\textit{\textbf{Enhanced Taylor diagrams using small multiple and interactive filtering.} The diagrams are utilized to visually compare seven Gaussian Process models with the ground truth. Only two models are highlighted for easier visual comparison.}}
    \label{fig: Small Multiple}
\end{figure*}

Hyper-parameter tuning is essential in optimizing ML models' performance. Tracking changes during tuning helps identify optimal parameter combinations, replicate successful experiments, avoid repeating unsuccessful ones, and analyze the effects of different hyper-parameter configurations on model performance. \autoref{fig: Small Multiple}, along with Supplemental Video 1, visualizes the data set inherited from a study by \textcite{ml-tuning-zewen}, designed for the regression task of predicting a sine function. It includes predictions generated by Gaussian Process models with varying hyper-parameters $\sigma_F$ and $\sigma_L$.
In order to closely inspect the dynamics of the data set, we interactively selected the models named \textit{BCM} and \textit{MoE}. Our analysis shows that both models perform best with $\sigma_F = 0.2$ and $\sigma_L = 0.5$. Additionally, we can see that small changes to both hyperparameters do not significantly affect the performance of the selected models, and they generally perform well without any anomalous predictions.

\section{Conclusion and Discussion}


We propose combining overview+detail and small multiple techniques with aggregation and interactive filtering for polar scatter charts, particularly in summary polar diagrams like the Taylor and mutual information diagrams. The examples in \autoref{sec: application examples} briefly illustrate the advantages of these design techniques over existing methods. Our user study and pretest review showed that these techniques enhanced comprehension and data exploration with summary polar diagrams, benefiting users regardless of their visualization expertise.\\
However, our study identified several limitations.
First, while we recognize the value of comparative studies, our research strategically prioritizes a rich, contextualized understanding of a single tool's impact on user tasks. This is especially important due to our tool being the first one that integrates summary polar diagrams combined with additional visualization techniques and the lack of directly comparable tools that integrate our specific combination of visualization techniques~\cite{user_study_argument_1}.
However, it is difficult to make claims about overall user performance without establishing a baseline using the previously published summary polar diagrams, which had limited interactivity.
Moreover, future work could explore evaluating the individual contributions of each visualization technique within our tool to better understand their specific impact on user performance.
Second, future research should examine larger sample sizes and additional factors affecting user performance, as the lack of significant differences may arise from small samples or high variability rather than true differences between experts and non-experts.
Third, when visualizing fewer than ten data points, the benefits of overview+detail and aggregation may diminish while demanding more screen space.
Fourth, existing high-level visualization libraries provide limited support for polar charts and interactive functionalities like brushing, which require further improvement.
Fifth, the proposed dashboard could also be more generalized and modularized for integration into other systems, extending the hybrid approach beyond summary polar diagrams to all scatter polar charts.
Sixth, we recommend limiting small multiples to a $3 \times 3$ grid maximum, as larger grids may reduce effectiveness~\cite{small-multiple-limitation}.
Seventh and last, the scalability of the proposed approach is constrained by human perceptual and cognitive limitations and current implementation. As highlighted in the TRS, the existing concept of enhanced summary polar diagrams cannot surpass the threshold of 21 distinct models (data set samples). While the wine data set described in~\autoref{sec: application examples} accommodates 20 samples, larger data sets are incompatible with current summary polar diagrams unless substantial modifications are implemented (\textit{e.g.,}~using the shape channel). Future research could focus on addressing these limitations.



\subsection*{Availability of data and materials}
All supplementary materials, source code, data sets, data generation scripts, and instructions are open-sourced and available at \url{https://github.com/AAnzel/Polar-Diagrams-Dashboard}, released under a GPL-3.0 license. Supplemental Video 1 can be found at the following link: \url{https://github.com/AAnzel/Polar-Diagrams-Dashboard/tree/master/data/Supplemental_Video_1}.

\subsection*{Author contributions}
A.A. contributed to conceptualization, data curation, investigation, validation, visualization, and writing. Z.Y. contributed to analysis and visualization. G.H. contributed to supervision, proofreading, and editing.


\subsection*{Acknowledgements}
We are thankful to the participants of our user study and the expert reviewers, whose valuable feedback was instrumental in shaping this research.
The authors further express their gratitude for the financial support granted by the Germany Federal Ministry of Health (BMG) under grant No. \texttt{2523DAT400} (project ``AI-assisted analysis and visualization of pandemic situations'', AI-DAVis-PANDEMICS).


\subsection*{Declaration of competing interest}

The authors have no competing interests to declare that are relevant to the
content of this article.

\printbibliography






\end{document}